\documentclass[10pt,aps,prd,twocolumn,floatfix,nofootinbib,showpacs,superscriptaddress]{revtex4-1}
\usepackage{graphicx,amsmath,amssymb,bm}
\usepackage{color}
\usepackage{amsmath}
\usepackage{amssymb}
\usepackage{graphicx}
\usepackage{multirow}
\usepackage{epsfig}
\usepackage{longtable} 
\usepackage{url}
\usepackage[utf8]{inputenc}
\usepackage{csquotes}

\usepackage{amsmath}
\usepackage{amssymb}
\usepackage{graphicx}
\usepackage{multirow}
\usepackage{epsfig}
\usepackage{rotating}
\usepackage{longtable} 
\usepackage{url}
\usepackage[utf8]{inputenc}
\usepackage{csquotes}
\usepackage{placeins}
\usepackage{hyperref}
\usepackage{scalerel}

\definecolor{purple}{rgb}{0.5,0,0.5}
\definecolor{blue}{rgb}{0.0,0,0.9}

\newcommand{\be}{\begin{equation}}
\newcommand{\ee}{\end{equation}}
\newcommand{\bea}{\begin{eqnarray}}
\newcommand{\eea}{\end{eqnarray}}
\newcommand{\ba}{\begin{array}}
\newcommand{\ea}{\end{array}}

\newcommand{\ovl}{\overline} 
\newcommand{\ie}{{\emph i.e.,\ }}

\newcommand{\ubar}{u^c} 
\newcommand{\dbar}{d^c} 
\newcommand{\ddbar}{D^c} 
\newcommand{\nubar}{\nu^c_e} 
\newcommand{\lbar}{\ovl{L}} 
\newcommand{\eplus}{e^+}

\newcommand{\spa}{\ }
\newcommand{\fac}{\eta_{331}}
\newcommand{\facd}{\eta_{LR}}
\newcommand{\fact}{\eta_{IN}}
\newcommand{\s}{M^0}

\begin{document}

\title{Flipped versions of the universal   3-3-1 and the  left-right symmetric models  in $[SU(3)]^3$:  a comprehensive approach}

 \author{Oscar Rodríguez}
 \email{oscara.rodriguez@udea.edu.co}
 \affiliation{Instituto  de F\'isica, Universidad de Antioquia,\\Calle 70 No.~52-21, Medell\'in, Colombia}
 \author{Richard H.~Benavides}
 \email{richardbenavides@itm.edu.co}
 \affiliation{Facultad de Ciencias Exactas y Aplicadas, Instituto Tecnológico Metropolitano,
 \\Calle 73 No 76 A - 354 , Vía el Volador, Medellín, Colombia}
 \author{William A.~Ponce}
 \email{wponce@gfif.udea.edu.co}
  \affiliation{Instituto  de F\'isica, Universidad de Antioquia,\\Calle 70 No.~52-21, Medell\'in, Colombia}
 \author{and Eduardo Rojas}
 \email{rojas@gfif.udea.edu.co}
 \affiliation{Instituto  de F\'isica, Universidad de Antioquia,\\Calle 70 No.~52-21, Medell\'in, Colombia}

\begin{abstract}
By  considering the  3-3-1 and the left-right symmetric  models as low  energy effective
  theories of the $SU(3)_C\otimes SU(3)_L\otimes SU(3)_R$~(for short  $[SU(3)]^3$) gauge group,  alternative versions  of these models are found.  
The new neutral gauge bosons of the universal 3-3-1 model and its flipped
versions are presented; also, the left-right symmetric model 
and its  flipped variants are  studied. 
Our analysis shows that there are  two  flipped versions of the universal 3-3-1 model,  
with the particularity that both  of them have the same weak charges.  
For the left-right symmetric model we also 
found two flipped versions; one of them  new in the 
literature which, unlike those of the 3-3-1, requires a dedicated study of its electroweak properties.
For all the models analyzed, the couplings of the $Z'$ bosons to the  standard model 
fermions are reported. The explicit form 
of the null space of the vector boson mass matrix 
for an arbitrary Higgs tensor and gauge group is also presented.
In the general framework of the  $[SU(3)]^3$ gauge group,
and by using the  LHC experimental results and EW precision data,
 limits on  the  $Z'$ mass  and the 
mixing angle between $Z$  and the new gauge bosons $Z'$ are 
obtained.
The general results call for very small mixing angles in
the range $10^{-3}$ radians and $M_{Z'}>$ 2.5 TeV.

\end{abstract}

\maketitle
\section{Introduction}
\label{sec:intro}
The quantization of the electric charge is an indication  
that the Standard Model~(SM) of the strong, weak and electromagnetic interactions
based on the local gauge group $SU(3)_C\otimes SU(2)_L\otimes U(1)_{Y}$,
might be embedded into a larger gauge structure~\cite{Pati:1974yy,Georgi:1974sy}.
This feature can be explained by grand unified theories~(GUT)
which, in general, have a unified coupling constant for all the 
interactions at  an energy given by  the GUT scale which is around 
$10^{16}$GeV for supersymmetric models.
One of the most important results of the 
GUT is the prediction of the neutrino masses in the $(10^{-5}-10^{2})$eV range~\cite{Langacker:1980js,Sayre:2006ma},
which is compatible with the present constraints on the neutrino masses~\cite{Agashe:2014kda}.

At the late seventies the unification theories were 
under suspicion owing to the prediction of topological
defects which are  typical  GUT predictions; 
from these considerations  the  cosmological 
inflation scenario  was born~\cite{Guth:1980zm}, 
which proved to be  quite useful to solve other  
cosmological problems, showing in this way 
that the insight provided by GUT is in the right direction.  
In general, the unification models  based on a simple group, 
 in particular the non-supersymmetric models, lead
to a detectable  proton decay~\cite{Langacker:1980js}.
However, when the group is the  product of two or more 
simple groups, the structure not necessarily contains gauge bosons that
mediate proton decay~\cite{Pati:1974yy,Willenbrock:2003ca,Ponce:1993ch}.
In this context, the trinification group based on the semisimple 
group $SU(3)_C\times SU(3)_L\times SU(3)_R$\footnote{ In trinification,      the equality  of the coupling constants at the   unification scale is assumed, 
which  is equivalent to impose an additional discrete $Z_3$ symmetry~(see \cite{Babu:1985gi} and references therein).  
In the present work 
such  assumption has not been made.}
~\cite{Achiman:1977py,Achiman:1978vg,Glashow:1984gc,Babu:1985gi},
results quite   convenient from a
phenomenological point of view  owing to the fact that  the 
baryon number is conserved by the gauge interactions~\cite{Willenbrock:2003ca}.
The original $SU(3)_L\times SU(3)_R$ models  with a lepton nonet
were first considered by Y. Achiman~\cite{Achiman:1977py,Achiman:1978vg};
however, earlier work on the $[SU(3)]^3$ group can be traced back up to the seminal works
in references~\cite{Salam:1964ry,Weinberg:1971nd}.
Besides, this  model has been  flexible enough to adjust  recent LHC anomalies, for example,  the di-photon excess 
at 750~GeV~\cite{diphoton}  and the di-boson excess 
at 1.9~TeV~\cite{Dias:2015mhm,Pelaggi:2015knk}.

The different  $[SU(3)]^3$ models have a rich phenomenology 
in the Higgs and neutrino sectors~\cite{Babu:1985gi,Hetzel:2015bla,Pelaggi:2015knk};
its  rank is  6~(equal to  $E_6$),
hence the model predicts, in addition to those already present in the SM,   
two additional heavy vector neutral gauge  bosons  which constitute one of the most important sources 
of constraints for the  model.  
In this paper we undertake  a detailed study 
of the couplings of these new gauge   bosons to the SM fermions, in order 
to put Electroweak~(EW) and collider constraints on $[SU(3)]^3$. \\

 In general, intricate models are not appealing. A way to look for new models with a moderate content of fermions  
is to consider flipped versions of the already  known  models in 
the literature~\cite{Barr:1981qv,Robinett:1982tq,Witten:1985xc,Ma:1986we,Ma:1995xk,Martinez:2001mu,Rojas:2015tqa,Mantilla:2016sew}.
An exhaustive account of the 
phenomenology of these models has not been done so far. Our work represents a first step in that  direction.
The first alternative model was ``Flipped $SU(5)$''~\cite{Barr:1981qv,Derendinger:1983aj},
which produces a symmetry breaking for $SO(10)$ GUT down to $SU(5)\otimes U(1)$, 
where the $U(1)$ factor contributes to the electric charge, 
and as such, its basic predictions for $\sin^2 \theta_W$  and the proton decay 
are known to be different from those of $SU(5)$. 
In the present work, we study the flipped versions
of the universal 3-3-1  and the left-right symmetric models   in $[SU(3)]^3$. 
That is equivalent to the  study of the 
  the different embeddings of  the SM fermions in the multiplets when the $[SU(3)]^3$ 
gauge group breaks  down  to the SM. 
 As a consequence of the reduction in the effective group symmetry,  
 these models predict  new $Z'$ bosons  at low energies.  
 For a given $Z'$ mass, these vector boson resonances 
 have well determined predictions in low energy experiments and colliders.
For universal models in the $E_6$ context, 
a systematic study of its  alternative  models  and further references can be found in~\cite{Rojas:2015tqa}.

The heavy vector bosons  $Z'$  are a generic prediction of the physics Beyond the SM~(BSM) 
with an  extended EW  sector~\cite{Langacker:2008yv}.  
The detection of one of these resonances at the LHC  will shed light on
the underlying symmetries  of the BSM physics. 
For the high luminosity regime
the  LHC will have  sensitivity for   $Z'$ masses under  5 TeV~\cite{Salazar:2015gxa,Godfrey:2013eta};
thus, a systematic and   exhaustive study  of the  EW extensions   
of the SM  with a minimal content of exotic fields  is mandatory.  
By imposing  universality on the  EW  extensions of the SM (as it happens in the SM),  
the possible EW extensions are basically the   $E_6$ subgroups~\cite{Slansky:1981yr,Carena:2004xs,Erler:2011ud}. 
$[SU(3)]^3$ is one of the four maximal $E_6$ subgroups; so, 
an exhaustive study of its neutral current structure is convenient,
something done in the present work.
As we will show, the couplings of additional gauge bosons to the SM fermions   are independent  of 
the Higgs sector and just depend on the $[SU(3)]^3$ symmetries. 
We also present LHC and EW constraints
for these models. 

 Finally, let us mention that  unification is not implicit in our assumptions; so, 
 non-universal gauge coupling strengths are used in this study.

The paper is organized as follows: 
in  Section~\ref{sec:trinification} we 
review the $[SU(3)]^3$ model and its subgroups. 
In Section~\ref{sec:331} we calculate the   EW couplings
for $Z'$ bosons in the $[SU(3)]^3$ subgroup $SU(3)_C\otimes SU(3)_L\times U(1)\otimes U(1)'$.
In Section~\ref{sec:higgs} we calculate  the eigenstates of the 
most general $[SU(3)]^3$ Higgs potential and, for considering different cases,  it is shown 
that these eigenstates are independent of the Higgs sector. 
It is also shown that the null space of the  
$[SU(3)]^3$ Higgs potential corresponds to the photon. 
In Section~\ref{sec:alr} we calculate the EW couplings 
for the left-right model and its alternative models. 
In Section~\ref{sec:ewconstraints} we impose EW 
and collider constraints  on the $Z\text{-}Z'$ mixing angle 
and  on the mass of the  new neutral $Z'$ gauge  bosons. 
Section~\ref{sec:conclusions} summarizes  our conclusions.
 Four  technical appendixes are presented at the end of the manuscript,  
in particular, in  Appendix~\ref{appendixb}  the null vector of  the EW vector boson mass matrix  is built  for  an arbitrary  
Higgs tensor  and gauge theory.
\section{The \texorpdfstring{$[SU(3)]^3$}{su31} group}
\label{sec:trinification}
The $[SU(3)]^3$ group~\cite{Slansky:1981yr,Georgi:1985,Babu:1985gi} 
$ SU(3)_C\otimes SU(3)_L\otimes SU(3)_R\equiv [SU(3)]^3$ is 
a maximal subgroup of $E_6$~\cite{Gursey:1975ki}  
with the same rank and fundamental representation. The three factor groups are identified in the following way: 
the first one corresponds to the 
vector like QCD color group $SU(3)_C$, the same as in the SM, and the other two can be identified with the 
left-right symmetric flavor 
group $SU(3)_L\otimes SU(3)_R$ extension of the $SU(2)_L\otimes SU(2)_R$, where $SU(2)_L$ in the SM is 
such that $SU(2)_L\subset SU(3)_L$. 
Using $\lambda_i,\;\; i=1,2,\dots ,8$ as the eight Gell-Mann matrices for $SU(3)$ normalized as 
$\text{Tr}(\lambda_i\lambda_j)=2\delta_{ij}$, the charge operator for the $[SU(3)]^3$ group may be written as 
\begin{equation}\label{charge}
 Q=\frac{\lambda_{3L}}{2} \oplus\frac{\lambda_{8L}}{2\sqrt{3}}\oplus\frac{\lambda_{3R}}{2} \oplus\frac{\lambda_{8R}}{2\sqrt{3}}\ .
\end{equation}
In this way, each family of fermions is assigned to a 27 as
\footnote{
Another convention  assigns leptons $\sim(1,\bar{3},3)$,
quarks $\sim (\bar{3},3,1)$ and antiquarks $\sim (3,1,\bar{3})$, 
in this case the assignments  of the $SU(3)_C$ representation of the quarks are interchanged 
with respect to the SM. 
In the present work we follow the Robinett and Rosner convention~\cite{Robinett:1982tq,London:1986dk}.}
\[27= (3,3,1)\oplus(1,\bar{3},3)\oplus(\bar{3},1,\bar{3})\ ,\]
where according to (\ref{charge}), the particle content of each term is:

\begin{eqnarray*}
(3,3,1)&=&(u,d,D)^T_L\ ,\\ 
(\bar{3},1,\bar{3})&=& (u^c,d^c,D^c)^T_L\ ,\\ 
(1,\bar{3},3)&=&\left(\begin{array}{ccc}
                       N^0 & E^- & e^-\\
                       E^+ & N^{0c} & \nu_e \\
                       e^+ & \nubar & M^0
                      \end{array}\right)_L,\end{eqnarray*}
which corresponds to the 27 states in the fundamental representation of $E_6$.

\subsection{3\text{-}3\text{-}1 models from \texorpdfstring{$[SU(3)]^3$}{SU32}}
\begin{widetext}

\begin{table}[t]  
\begin{center}
\scalebox{0.8}{
\begin{tabular}{|c|c|c|}
\hline
$U(1)^\prime$\cite{Rojas:2015tqa}  &$Q'$&  Charges\\
\hline
$U_{R}$& 2$I_{R3}$  
& $(\eplus,\dbar,\lbar)_{+1} +(l,q,D,\ddbar,\s)_{0} + (\nubar,\ubar,L)_{-1}$ \\
$U_{I}$& 2$U_{R3}$  
& $(\nubar,\ddbar,L)_{+1} +(\lbar,q,\ubar,D,\eplus)_{0}+(\s,\dbar,l)_{-1}$ \\
$U_{A}$& 2$V_{R3}$   
 & $(\s,\ubar,l)_{+1} + (L,q,\dbar,D,\nubar)_{0}+(\eplus,\ddbar,\lbar)_{-1}$ \\
\hline
 $U_{33}$& $2\sqrt{3}I_{L8}$    
&  $(l,\lbar,L)_{-1} +(\ubar,\dbar,\ddbar)_{0} + (\eplus,\nubar,\s)_{+2}+q _{+1}+D_{-2} $ \\
 $U_{21\ovl{R}}$  & $-2\sqrt{3}I_{R8}$
&  $(\eplus,\nubar,\lbar,L)_{-1}+ (q,D)_{0} +(l,\s)_{+2} +(\ubar,\dbar)_{+1}
+\ddbar_{-2}$ \\
 $U_{21\ovl{I}}$  & $-2\sqrt{3}U_{R8}$
&  $(\s,\nubar,l,L)_{-1} + (q,D)_{0} +(\lbar,\eplus) _{+2}+(\ddbar,\dbar)_{+1}+
\ubar_{-2}$ \\
 $U_{21\ovl{A}}$ & $-2\sqrt{3}V_{R8}$
&  $(\s,\eplus,l,\lbar)_{-1} + (q,D)_{0} +(L,\nubar)_{+2}+(\ddbar,\ubar)_{+1}+
\dbar_{-2}$ \\
\hline 
$U(1)_{31R}$ & $I_{BL} $ &  $(\lbar,L,\s)_{0}      + q_{+1/6} +(\ubar,\dbar)_{-1/6} +(\eplus,\nubar)_{+1/2}+l_{-1/2}    +\ddbar_{+1/3}+D_{-1/3}$ \\
$U(1)_{31I}$ & $U_{BL}$  &  $(l,L,\eplus)_{0}     + q_{+1/6} +(\ddbar,\dbar)_{-1/6}+(\s,\nubar)_{+1/2}     +\lbar_{-1/2}+\ubar_{+1/3} +D_{-1/3}$ \\
$U(1)_{31A}$ & $V_{BL}$  &  $(l,\lbar,\nubar)_{0} + q_{+1/6} +(\ddbar,\ubar)_{-1/6}+(\s,\eplus)_{+1/2}     +L_{-1/2}    +\dbar_{+1/3} +D_{-1/3}$ \\ 
\hline
\end{tabular}} 
\caption{Charge assignments for the fundamental 
representation of the  $[SU(3)]^3$ group,  the same  27 of the   $E_6$ group, under different $U(1)$ symmetries.  
 For the first family, $l$  is the SM lepton doublet,  $l^T=(\nu, e^{-})$ and $q=(u,d)^T$ is the   SM quark doublet.
The  charge conjugated  of the corresponding right handed weak-isospin singlets
are $e^c$,  $\nu^c$, $u^c$  and  $d^c$. 
The heavy exotic particles are vector under the SM group, the heavy down quark,  
$D$ ~( $D^c$), is an weak-isospin singlet (charge conjugated of the corresponding right handed
chiral projection) of charge $-1/3$~($+1/3$), $L=(N^0,E^{-})^T$ and $L=(E^{c},N^{0c})^T$, 
are additional weak-isospin doublets where  $L$ have the the same quantum 
numbers of the SM lepton doublet,  and  $\s$ is a singlet under 
 the SM.    }
\label{tab:e6charges}
\end{center}
\end{table}

\end{widetext}

Let us now consider the decomposition of the $[SU(3)]^3$ gauge group into a subgroup 
$G$ which survives at an intermediate energy scale between the EW scale (245 GeV) 
and the unification scale; that is $[SU(3)]^3\supset G$.

Suppose first that $G$ corresponds to the universal 3-3-1 model~\cite{Sanchez:2001ua} 
\begin{align}
 G&=SU(3)_C\otimes SU(3)_L\otimes U(1)_X\notag\\
  &\subset SU(3)_C\otimes SU(3)_L\otimes U(1)_a\otimes U(1)_b\ .
\end{align}
By using that $SU(3)\rightarrow SU(2)_a\otimes U(1)_b$ 
the triplet in each   nonet  goes to a doublet with charge $b$ and a singlet with 
charge $-2b$, \ie $3\rightarrow 2_{b}+1_{-2b}$.  
Next by breaking    the remaining spin symmetry, \ie $SU(2)_a\rightarrow U(1)_a$,  the
doublet goes to a couple of singlets, \ie  $2_{b}+1_{-2b}\rightarrow 1_{a,b}+1_{-a,b}+1_{0,-2b}$.
Thus, when $SU(3)_R$ breaks into $U(1)_a\otimes U(1)_b$ the following branching rule applies: 
\begin{equation}\label{br3}
3_R\longrightarrow (a)(b)+(-a)(b)+(0)(-2b)\ ,
\end{equation}
which implies:
\begin{eqnarray*}
 (3,3,1)&\longrightarrow& (3,3,0,0)\ ,\\
 (\bar{3},1,\bar{3})&\longrightarrow& (\bar{3},1,-a,-b)\oplus (\bar{3},1,a,-b)\oplus (\bar{3},1,0,2b)\ ,\\
 (1,\bar{3},3)&\longrightarrow& (1,\bar{3},a,b)\oplus (1,\bar{3},-a,b)\oplus (1,\bar{3},0,-2b)\ ;
\end{eqnarray*}
because  the nonet   $(3,3,1)$ is simultaneously  a  color and a $SU(3)_L$ triplet,   the unique possibility for 
the fermion assignment is
\begin{align*}
 (3,3,1)\longrightarrow &(3,3,0,0) 
              =   (u_L,d_L,D_L)^T_{0}\ .
\end{align*}
For the nonet $(\bar{3},1,\bar{3})$  there are three different fermion assignments
in consistency with  the three 
different $SU(2)_X$ spin symmetries\footnote{In Appendix \ref{sec:su3sctruture} we briefly  review the $SU(2)$ weak-I-spin~(or Isospin), weak-$U$-spin  and weak-$V$-spin  symmetries 
in $SU(3)$.}~\cite{London:1986dk}, $X= I,U$ and $V$, \ie
\begin{align*}
 &(\bar{3},1,\bar{3})\longrightarrow (\bar{3},1,-a,-b)\oplus (\bar{3},1,a,-b)\oplus (\bar{3},1,0,2b)\\
               &
 =
 \begin{cases}
 (d_L^c)_{-a,-b}\oplus (u_L^c)_{a,-b}\oplus (D_L^c)_{0,2 b}\ ,\hspace{0.5cm}X=I\ ,\\
 (D_L^c)_{-a,-b}\oplus (d_L^c)_{a,-b}\oplus (u_L^c)_{0,2 b}\ ,\hspace{0.5cm}X=U\ ,\\
 (u_L^c)_{-a,-b}\oplus (D_L^c)_{a,-b}\oplus (d_L^c)_{0,2 b}\ ,\hspace{0.5cm}X=V\ .
 \end{cases}
 \\
\end{align*}
We label the three possible fermion assignments with 
$X=I,U,V$, which denote weak-$I$-spin, weak-$U$-spin and weak-$V$-spin, respectively.
As can be seen, the $[SU(3)]^3$ gauge group produces three different low energy 3-3-1 fermion 
structures; the ordinary one presented in reference~\cite{Sanchez:2001ua}, and two more new  in 
the literature as far as we know.

In a corresponding way, there are three different fermion assignments for the nonet $(1,\bar{3},3)$,
\ie 
\begin{align*}
 &(1,\bar{3},3)\longrightarrow (1,\bar{3},a,b)\oplus (1,\bar{3},-a,b)\oplus (1,\bar{3},0,-2b)\\
              &=
\begin{cases}
  (E^-_L,N^{0c}_L,\nu^c_{eL})^T_{a,b} \oplus (N^0_L,E^+_L,e^+_L)^T_{-a,b}\notag\\      
 \hspace{0.5cm}   \oplus (e^-_L,\nu_{eL},M^0_L)^T_{0,-2b}\ ,\hspace{0.5cm}X=I\ ,\\
  (e^-_L,\nu_{eL},M^0_L)^T_{a,b}      \oplus (E^-_L,N^{0c}_L,\nu^c_{eL})^T_{-a,b}\notag\\  
 \hspace{0.5cm}  \oplus (N^0_L,E^+_L,e^+_L)^T_{0,-2b}\ ,\hspace{0.5cm}X=U\ ,\\
  (N^0_L,E^+_L,e^+_L)^T_{a,b}         \oplus (e^-_L,\nu_{eL},M^0_L)^T_{-a,b} \notag\\    
 \hspace{0.5cm}    \oplus (E^-_L,N^{0c}_L,\nu^c_{eL})^T_{0,-2b}\ ,\hspace{0.5cm}X=V\ .\\
\end{cases}              
\end{align*}
In correspondence with Eq.~(\ref{charge}), the electric charge is now given by
\begin{equation}\label{chargeab}
 Q=I_{L3} +\frac{1}{\sqrt{3}}I_{L8}+c_X X_{R3} +\frac{2d_X}{\sqrt{3}} X_{R8}\ ,
\end{equation}
where  $X_{R3}$ and $X_{R8}$ are the fermion charges under $U(1)_a$ and $U(1)_b$, respectively,
as it is shown in Table~\ref{tab:e6charges}, and
$c_X$ and $d_X$ are
\begin{align*}
 c_I =&\quad\;  1\ ,\hspace{1cm}d_I=1/2\ ,\\
 c_U =&\quad\;  0\ ,\hspace{1cm}d_U=-1 \ ,\\
 c_V =&        -1\ ,\hspace{1cm}d_V=1/2\ ,
\end{align*}
where we have taken   $b=1/(2\sqrt{3})$ and $a=1/2$ in order to have the 
charges properly normalized as in  $E_6$.
In Eq.~(\ref{chargeab})  $I_{L3}$ and $I_{L8}$  represent  the  charges of the  fermions in the  $27$, 
when these operators  act on the triplets;  in the nonets the 
  corresponding tridimensional representation are   $\lambda_{3L}/2$ and 
$\lambda_{8L}/2$,  respectively~[see  Eq.~(\ref{charge})].
In the same vein in  Eq.~(\ref{chargeab}) with $X=I$, 
the charges  $I_{R3}$ and $I_{R8}$ correspond to 
$\lambda_{3R}/2$ and  $\lambda_{8R}/2$, respectively.
The difference between the weak-$U$-spin and the alternative  3-3-1 models  (the normal and  the flipped one)
is the  interchange of fermions between the multiplets, 
something which does not affect the low energy phenomenology for the neutral sector as we will 
see in the next Section.\\

\subsection{Left-right symmetric models from \texorpdfstring{$[SU(3)]^3$}{su34}}
A further step is to take $G=SU(3)_C\otimes SU(2)_L\otimes SU(2)_X\otimes U(1)_f\otimes U(1)_g$, 
which is obtained by using the branching rule for $SU(3)_{L,R}\longrightarrow SU(2)_{L,X}\otimes U(1)_{f,g}$ as 
\begin{equation*}\label{br32}
3_L\longrightarrow (2, f) + (1,-2f)\ ,\hspace{0.5cm}
3_R\longrightarrow (2, g) + (1,-2g)\ ,
\end{equation*}
which produces three different ways to reach the $U(1)_Y$ in the SM
\begin{eqnarray*}
(3,3,1)&\longrightarrow& (3,2,1,f,0)\oplus (3,1,1,-2f,0)\ ,\\
(\bar{3},1,\bar{3})&\longrightarrow& (\bar{3},1,\bar{2},0,-g))\oplus (\bar{3},1,1,0,2g)\ ,\\
(1,\bar{3},3)&\longrightarrow& (1,\bar{2},2,-f,g)\oplus (1,\bar{2},1,-f,-2g)\notag\\
&&\oplus (1,1,2,2f,g)\oplus~(1,1,1,2f,-2g)\ .
\end{eqnarray*}
The underlying breaking behind these branching rules are:
\begin{align*}
 (3,3,1)            \longrightarrow & (3,2_{f},1_{0})\oplus (3,1_{-2f},1_{0})\ ,\notag\\
 (\bar{3},1,\bar{3})\longrightarrow &(\bar{3},1_{0},\bar{2}_{-g})\oplus (\bar{3},1_{0},1_{2g})\ ,\notag\\
(1,\bar{3},3)       \longrightarrow &(1,\bar{2}_{-f},2_{g})\oplus (1,\bar{2}_{-f},1_{-2g})\notag\\
              &\oplus (1,1_{2f},2_{g})\oplus (1,1_{-2f},1_{2g})\ .
\end{align*}
Now the definition of   $U(1)_{BLX}\equiv U(1)_f+U(1)_g$ for $f=g=1/6$  conducts to  the alternative left-right symmetric models 
\[SU(3)_C\otimes SU(2)_L\otimes SU(2)_X\otimes U(1)_{BLX}\ ,\]
with the following particle content for the quark sector:

\begin{align*}
 (3,3,1) =&(u,d,D)_L\longrightarrow (3,2,1,1/6)\oplus (3,1,1,-1/3) \notag\\
         =&(u,d)_L\oplus D_L\ ,\\  
 (\bar{3},1,\bar{3})=& (u^c,d^c,D^c)_L\longrightarrow (\bar{3},1,\bar{2},-1/6)\oplus (\bar{3},1,1,1/3)\\
              =&
\begin{cases}
(u^c,d^c)_L\oplus D_L^c\ ,\hspace{1.5cm}X=I\ ,\\
(D^c,d^c)_L\oplus u_L^c\ ,\hspace{1.5cm}X=U\ ,\\
(u^c,D^c)_L\oplus d_L^c\ ,\hspace{1.5cm}X=V\ .
\end{cases}
\end{align*}
For the lepton sector we have:
\begin{align*}
&(1,\bar{3},3)=\left(\begin{array}{ccc}
                       N^0 & E^- & e^-\\
                       E^+ & N^{0c} & \nu_e \\
                       e^+ & \nu^c_e & M^0
                      \end{array}\right)_L\\
                      \longrightarrow & (1,\bar{2},2,0)\oplus (1,\bar{2},1,-1/2)\oplus (1,1,2,1/2)
\oplus (1,1,1,0)\\
&=
\begin{cases}
    \left(\begin{array}{cc}
                       E^+ & N^{0c}\\
                       N^0 & E^-\\
                      \end{array}\right)_L\oplus 
    \left(\begin{array}{c}\nu_e\\e^-
                       \end{array}\right)_L\oplus (e^+, \nu^c_e)_L \\
                     \hspace{4.2cm}\oplus\  M^0_L\ ,\hspace{0.5cm}X=I\ ,
\\
  \left(\begin{array}{cc}
                       \nu_e & N^{0c}\\
                       e^-   & E^-\\
                       \end{array}\right)_L\oplus 
     \left(\begin{array}{c}E^+\\ N^0
                        \end{array}\right)_L\oplus 
                        (M^0_L, \nu^c_e)_L\\
                        \hspace{4.2cm}\oplus\  e^+\ ,\hspace{0.5cm}X=U\ ,  
 \\                      	
  \left(\begin{array}{cc}
                        E^+ & \nu_e\\
                        N^0 & e^-\\
                       \end{array}\right)_L\oplus 
     \left(\begin{array}{c}N^{0c}\\ E^-
                        \end{array}\right)_L\oplus 
                        (M^0_L, e^+)_L\\
                        \hspace{4.2cm}\oplus\  \nu^c_e\ ,\hspace{0.5cm}X=V\ ,                          
\end{cases}                        
\end{align*}
In the left-right model 
$SU(2)_L\otimes SU(2)_R\otimes U(1)_{B-L}\subset SU(3)_L\otimes SU(3)_R$~(in our notation $SU(2)_L\otimes SU(2)_I\otimes U(1)_{BLI}$ )
the weak-isospin subgroup~($X=I$) has been used. 
That is the correct choice for the left-handed sector, but not the only choice for 
the right-handed one as we have shown  already. The weak-$V$-spin  symmetric case is 
a very well known  example where    $SU(2)_V$ is used  instead of  $SU(2)_R$; this model 
is known  as the  Alternative left-right\footnote{Or alternate left-right Model.}(ALR), 
which was found in a different way in Ref.~\cite{Ma:1986we}. The case $X=U$ is a new  alternative model. 
\section{3-3-1 Neutral currents}
\label{sec:331}
For the $[SU(3)]^3$ group the interaction   Lagrangian $-\mathcal{L}_I$ is
\begin{align}\label{eq:current}
 &g_L J_{L3\mu}^{I}A_{L3}^{I\mu}+g_L J_{L8\mu}^{I}A_{L8}^{I\mu}+g_R J_{R3\mu}^{X}A_{R3}^{X\mu}+g_R J_{R8\mu}^{X}A_{R8}^{X\mu}\notag\\
=&g_L J_{L3\mu}^{I}A_{L3}^{I\mu}+g^{\prime}J_{Y\mu} B^\mu +g_2J_{2\mu} Z^{\prime\mu}+g_3 J_{3\mu} Z^{\prime\prime\mu}\ .
\end{align}
where $A_{L3\mu}^I, A_{L8\mu}^I, A_{R3\mu}^X$ and $A_{R8\mu}^X$ are the corresponding vector gauge bosons associated with 
$\lambda_{L3}^I, \lambda_{L8}^I, \lambda_{R3}^X$ and $\lambda_{R8}^X $, respectively~(for a  precise definition see Appendix~\ref{appendixa}).
The neutral  currents in (\ref{eq:current}) are given by 
\begin{align}
       J^X_{R8\mu}=& \sum_i \bar{f}_i \gamma_\mu [\epsilon_\text{{\bf L}}^{X_{R8}}(i)P_\text{{\bf L}}+\epsilon_\text{{\bf R}}^{X_{R8}}(i)P_\text{{\bf R}}]f_i\ , \\
       J^X_{R3\mu}=& \sum_i \bar{f}_i \gamma_\mu [\epsilon_\text{{\bf L}}^{X_{R3}}(i)P_\text{{\bf L}}+\epsilon_\text{{\bf R}}^{X_{R3}}(i)P_\text{{\bf R}}]f_i\ , 
\end{align}
where the chiral charges, $\epsilon_\text{{\bf L, R}}$, are shown in Table~\ref{tabla:1} in Appendix~\ref{appendixc1}. 
Notice in our notation   that the bold labels $\text{{\bf L, R}}$ refer to 
the left and right chiral  projections and $L$ and $R$ refer to different  $SU(n)$  group structures. 
By means of an orthogonal matrix  we can rotate from the $[SU(3)]^3$ 
basis of the neutral vector  bosons, to a basis where one boson corresponds to the hypercharge, \ie 
\begin{align}\label{eq:orthogonal}
\begin{pmatrix}
 A_{L3\mu}^I \\
 B_\mu \\
 Z'_\mu \\
 Z''_\mu 
\end{pmatrix}
= 
\mathcal{O}^{T}
\begin{pmatrix}
 A_{L3\mu}^I \\
 A_{L8\mu}^I \\
 A_{R8\mu}^X \\
 A_{R3\mu}^X \\
\end{pmatrix}
\ ,
\end{align}
where the orthogonal matrix  is 
\begin{align}
\mathcal{O}= 
\begin{pmatrix}
1 &0 &0  &0 \\
0 &1 &0  &0 \\
0 &0 &\cos \beta & -\sin \beta \\
0 &0 &\sin \beta &  \cos \beta
\end{pmatrix}
\begin{pmatrix}
1 &0           &0             & 0 \\
0 &\cos \alpha & -\sin \alpha & 0 \\
0 &\sin \alpha &  \cos \alpha & 0 \\
0 &0           &    0         & 1
\end{pmatrix}
\ .
\end{align}
It is important to realize that in order to recover the particular case $X=U$, corresponding to the 3-3-1 models,
it is necessary to take  $\cos \beta =-1$.  By replacing this expression in Eq.~(\ref{eq:current}) we obtain
\begin{align}
&g'B_\mu J_Y^\mu= B^\mu\Big(
 g_L J^I_{L8\mu} \cos \alpha \notag\\
&+g_R J^X_{R8\mu} \sin \alpha \cos \beta  
+g_R J^X_{R3\mu} \sin \alpha \sin \beta 
   \Big)\ ;
\end{align}
by equating with 
\begin{align}
J_{Y\mu}= \frac{1}{\sqrt{3}}J_{L8\mu}^I +c_XJ^{X}_{R3\mu}+\frac{2d_X}{\sqrt{3}}J_{R8\mu}^X\ ,  
\end{align}
we get the following three equations: 
\begin{align}\label{eq:eq331}
   \frac{1}{\sqrt{3}}g' =& g_L \cos \alpha\ ,\hspace{0.5cm}   
\frac{2d_X}{\sqrt{3}}g' = g_R \sin \alpha \cos \beta\ ,\hspace{0.5cm} \\
                c_X g' =&  g_R  \sin \alpha \sin \beta\ .
\end{align}
From these equations we have, 
\begin{align}
 g_R=& \sqrt{\frac{N}{F}}g' 
\ ,\hspace{1cm}
\cos \alpha =
\frac{g'}{\sqrt{3}g_L } 
\ , \notag\\        
\cos \beta  =& \frac{2d_X}{\sqrt{N}}=d_X\ ,
\end{align}
where $N= (3c^2_X +4 d^2_X)=4$,  and $F=3 - (g'/g_L)^2$.
It is worth to notice that in the three cases considered, \ie for any value of  $X$,
\begin{align}\label{eq:gr}
g_{R}= \frac{2 g_L g'}{\sqrt{3 g_L^2-g'^2}}
    =\frac{2g_L\sin \theta_W}{\sqrt{4\cos^2\theta_W-1}}\ ,
\end{align}
From the equations (\ref{eq:current})  and (\ref{eq:orthogonal}) it is possible to get expressions for 
the neutral currents associated with the  $Z'$ and $Z''$ bosons, respectively
\begin{align}\label{eq:currents}
g_2J_{2\mu}=& -g_L J^I_{L8\mu}  \sin \alpha + g_R J^X_{R8\mu} \cos \alpha \cos \beta \notag\\ 
            & +  g_R J^X_{R3\mu}\cos \alpha  \sin \beta\ ,\notag\\
g_3J_{3\mu}=&- g_R J^X_{R8\mu}  \sin \beta  + g_R J^X_{R3\mu} \cos \beta\ . 
\end{align}
From these relations and from Table~\ref{tabla:1}  we can obtain the  explicit expressions of the vector and axial charges
for the $Z'$ and  $Z''$ gauge bosons, these charges are shown in Tables~\ref{tab:331g} and \ref{tab:inertmodel}, respectively.
The collider an EW constraints are shown in Table~\ref{tab:limits} and Figure~\ref{Contours}.
A detailed analysis of these constraints is presented in Section~\ref{sec:ewconstraints}. 
Finally, we can make  use of the defining condition of the orthogonal matrices,   $\mathcal{O}^{-1}=\mathcal{O}^{T}$,
and use the matrix~(\ref{eq:orthogonal}) to    rotate from the $[SU(3)]^3$  basis
for the neutral vector bosons  to the SM basis, \ie 
\begin{align}\label{eq:orthogonal3}
&
\begin{pmatrix}
 A_{\mu} \\
 Z_\mu \\
 Z'_\mu \\
 Z''_\mu 
\end{pmatrix}
= \mathcal{W}\cdot\mathcal{O}^T
\begin{pmatrix}
 A_{L3\mu}^I \\
 A_{L8\mu}^I \\
 A_{R8\mu}^X \\
 A_{R3\mu}^X \\
\end{pmatrix}
\notag\\
=&
\begin{pmatrix}
\sin \theta_W  &  \cos \theta_W& 0         & 0 \\
\cos \theta_W  & -\sin \theta_W& 0         & 0 \\
0              &0             & 1         & 0 \\
0              &0             & 0         & 1
\end{pmatrix}
\mathcal{O}^T
\begin{pmatrix}
 A_{L3\mu}^I \\
 A_{L8\mu}^I \\
 A_{R8\mu}^X \\
 A_{R3\mu}^X \\
\end{pmatrix}
\ ,
\end{align}
where $\mathcal{W}$  and  $ \theta_W$ are the Weinberg matrix and the Weinberg angle, respectively.

\section{Eigenstates of the vector boson mass matrix in \texorpdfstring{$[SU(3)]^3$}{su35}}
\label{sec:higgs}
In the last section we saw that it is possible to  obtain the SM fields $A_{\mu}$ and $Z_{\mu}$ and the extra neutral vector bosons 
 $Z_{\mu}^{\prime}$  and   $Z_{\mu}^{\prime\prime}$ 
by rotating the $[SU(3)]^3$ basis for the vector fields. 
By making use of some viable cases for the Higgs potential
in the present section, we will show that, 
independent of the Higgs sector, the
null space of the vector boson mass matrix 
corresponds to the photon,  \ie  by rotating 
the photon component $(A_{\mu},0,0,0)^T$  in the SM basis 
to the $[SU(3)]^3$ basis, we obtain 
the null space of the vector boson mass matrix.
This is a  particular example of a more general theorem which 
is shown in Appendix~\ref{appendixb}. 
In that sense, the present section is useful to provide a context for this demonstration.
The same is not true for the  eigenvalues of the 
vector mass matrix which strongly depend on the Higgs sector. 
In the fundamental representation of the $[SU(3)]^3$ group 
the neutral components are in the leptonic sector $(1,\overline{3},3)$;
if we put the Higgs field $\Phi$ in the same representation 
the corresponding transformation properties are 
\begin{align}\label{eq:transph}
\Phi^\prime = U_L\Phi U_R^\dagger,\hspace{0.5cm}U_{L,R}= \exp\left(-i\theta^a(x)\lambda^a_{L,R}/2\right)\ . 
\end{align}
Requiring   gauge invariance,   the  covariant derivative is
\begin{align}
 D_\mu\Phi = \partial_\mu\Phi-\frac{i}{2}\left(g_L\lambda^a A^a_{\mu L}\Phi-g_R\Phi\lambda^a A^a_{\mu R}  \right)\ ,
\end{align}
which transforms in the same way as the Higgs fields, \emph{i.e.},
\begin{align}\label{eq:transDph}
(D_\mu\Phi)^\prime = U_L D_\mu\Phi U_R^\dagger\ , 
\end{align}
as it is required  to build the gauge invariant kinetic term. 
The Higgs sector of the $[SU(3)]^3$ model  contains two complex scalar field nonets, $\Phi_1$ and $\Phi_2$.
The most general  vacuum expectation value~(VEV) for these fields are~\cite{Hetzel:2015bla}
\begin{align}
 \langle \Phi_1\rangle= 
 \begin{pmatrix}
 v_1 & 0        & 0\\
 0 & b_1 & 0 \\
 0 &  0 & M_I
 \end{pmatrix},\hspace{0.2cm}
\langle \Phi_2\rangle=
 \begin{pmatrix}
 v_2 & 0        & 0\\
 0 & b_2 & b_3 \\
 0 &  M_R & M_3
 \end{pmatrix}
 \ ,
\end{align}
where $\Phi_1$ is diagonal in view  of  the fact that  it is always possible to bring one Higgs VEV into its diagonal form 
by using the $SU(3)_L\times SU(3)_R$ symmetry. 
The vector boson masses came from 
\begin{align}
\mathcal{L}_K
= 
+\sum_{i=1,2}\text{Tr}\left[D_\mu \Phi_i (D^\mu \Phi_i)^\dagger\right]|_{\Phi_i=\langle\Phi_{i} \rangle}\ ,
\end{align}
which is invariant under the already mentioned gauge transformations [Eqs.~(\ref{eq:transph}) and  (\ref{eq:transDph})].
We can get rid of
the kinetic mixing term $\text{Tr}\left[D_\mu \Phi_1 (D^\mu \Phi_2)^\dagger\right]$
by redefining the scalar fields  in order to cast the Lagrangian into the canonical form. 
By a rotation  in the adjoint representation we obtain 
the simplified expression 
\begin{align}   
&\mathcal{L}_K(\langle\Phi_{1} \rangle,\langle\Phi_{2} \rangle)= \frac{1}{3}\Bigg(\notag\\
         & +\left(g_L A_{L8\mu}^V +g_R A_{R8\mu}^I\right)^2b^2_3\notag\\
         & +\left(g_R A_{R8\mu}^V +g_L A_{L8\mu}^I\right)^2M^2_R \notag\\     
         & +\left(g_L A_{L8\mu}^V- g_R A_{R8\mu}^V\right)^2(b^2_1+b^2_2) \notag\\
         & +\left(g_L A_{L8\mu}^I -g_R A_{R8\mu}^I\right)^2(M_3^2+M_I^2) \notag\\
         & +\left(g_L A_{L8\mu}^U -g_R A_{R8\mu}^U\right)^2(v_1^2+v_2^2)
       \Bigg)\ ,   
\end{align}   
where 
$A_{(L,R)8\mu}^V=-(A_{(L,R)8\mu}^I -\sqrt{3}A_{(L,R)3\mu}^I)/2$ and
$A_{(L,R)8\mu}^U=-(A_{(L,R)8\mu}^I +\sqrt{3}A_{(L,R)3\mu}^I)/2$.  
By writing the kinetic part  in terms of $A_{R8\mu}^I$
and $A_{R3\mu}^I$, the Higgs covariant derivative can be written as	
\begin{align}
\mathcal{L}_K=\frac{1}{2}
\mathcal{A}^T\cdot\mathcal{M}\cdot\mathcal{A}
\ ,
\end{align}
where $\mathcal{A}=(A_{L3\mu}^I  A_{L8\mu}^I  A_{R8\mu}^I  A_{R3\mu}^I)^T$,
and  $\mathcal{M}$ is the gauge boson mass matrix whose   elements are given by   
\begin{align*}
 M_{11}=& \frac{2}{4}g_L^2(b_1^2+b_2^2+b_3^2+v_1^2+v_2^2)\ , \\ 
 M_{12}=& \frac{-2}{4\sqrt{3}}g_L^2 (b_1^2+b_2^2+b_3^2-v_1^2-v_2^2)\ ,   \\  
 M_{13}=& \frac{2}{4\sqrt{3}}  g_Lg_R(b_1^2+b_2^2-2b_3^2-v_1^2-v_2^2)\ ,\\ 
 M_{14}=& \frac{-2}{4}g_Lg_R(b_1^2+b_2^2+v_1^2+v_2^2)\ ,                 \\   
 M_{24}=& \frac{2}{4\sqrt{3}}  g_Lg_R(b_1^2+b_2^2-2M_R^2-v_1^2-v_2^2)\ , \\  
 M_{34}=& \frac{-2}{4\sqrt{3}}g_R^2 (b_1^2+b_2^2+M_R^2-v_1^2-v_2^2)\ ,    \\
 M_{22}=& \frac{2}{12}g_L^2(b_1^2+b_2^2+b_3^2+4M_3^2+4M_I^2\\
        &+4M_R^2+v_1^2+v_2^2)\ ,    \\
 M_{33}=& \frac{2}{12}g_R^2(b_1^2+b_2^2+4b_3^2+4M_3^2+4M_I^2\\
        &+M_R^2+v_1^2+v_2^2)\ ,    \\  
 M_{23}=& \frac{-2}{12}g_Lg_R(b_1^2+b_2^2-2b_3^2+4M_3^2+4M_I^2\\
        &-2M_R^2+v_1^2+v_2^2)\ , \\  
 M_{44}=& \frac{2}{4}g_R^2(b_1^2+b_2^2+M_R^2+v_1^2+v_2^2)\ .
\end{align*}
The null space of the mass matrix $\mathcal{M}$ is 
\begin{align}
\mathcal{A}_{\text{null}}^{\mu}=
\mathcal{N}
\begin{pmatrix}
\frac{1}{g_L}, & \frac{1}{\sqrt{3}g_L},  & \frac{1}{\sqrt{3}g_R},& \frac{1}{g_R}
\end{pmatrix}
A^{\mu}(x) \ ,
\end{align}
where $A^{\mu}(x)$ is an arbitrary vector field which, as we will see later, corresponds to 
the photon, and $\mathcal{N}$ is an arbitrary normalization.  
We can obtain a similar expression for the null eigenvector 
by  inverting   Eq.~(\ref{eq:orthogonal3})
\begin{align}\label{eq:eigenvector1}
\mathcal{R}
\begin{pmatrix}
A_{\mu}\\
0\\
0\\
0
\end{pmatrix}
=&g_L\sin \theta_W  
\begin{pmatrix}
\frac{1}{g_L} \\
\frac{1}{\sqrt{3}g_L} \\
\frac{1}{\sqrt{3}g_R} \\
\frac{1}{g_R}    
\end{pmatrix}
A_{\mu}
\ ,\notag\\
\hspace{0.5cm}
\mathcal{R}
\begin{pmatrix}
 0\\
 Z_\mu\\
 0\\
 0
\end{pmatrix}
=&
g_L \frac{\sin^2 \theta_W}{ \cos \theta_W}  
\begin{pmatrix}
\frac{\cot^2 \theta_W}{g_L} \\
-\frac{1}{\sqrt{3}g_L} \\
-\frac{1}{\sqrt{3}g_R} \\
-\frac{1}{g_R} 
\end{pmatrix} 
Z_\mu \ ,
\hspace{0.5cm}
\end{align}
where $\mathcal{R}=\left(\mathcal{W}\cdot\mathcal{O}^T\right)^T$.
In order to get this result it was necessary to impose
$g'= g_L \tan \theta_W $, which is also satisfied in the SM. 
This shows that the null vector corresponds 
to the photon as we previously said. 
By proceeding  in a similar way for   $Z^{\prime}_\mu$  and  
$Z^{\prime\prime}_\mu$ we find
\begin{align}\label{eq:eigenvector2}
\mathcal{R}
\begin{pmatrix}
 0\\
 0\\
 Z_\mu^\prime\\
 0
\end{pmatrix}
=&
\frac{g_L}{2} \tan \theta_W  
\begin{pmatrix}
   0 \\
-\frac{4}{\sqrt{3}g_R} \\
 \frac{1}{\sqrt{3}g_L} \\
 \frac{1}{g_L} 
\end{pmatrix} 
Z_\mu^\prime\ ,\notag\\
\hspace{0.5cm}
\mathcal{R}
\begin{pmatrix}
 0\\
 0 \\
 0\\
 Z_\mu^{\prime\prime}
\end{pmatrix}
=&
\begin{pmatrix}
0\\
0\\
-\sqrt{3}/2\\
1/2
\end{pmatrix} 
Z_\mu^{\prime\prime}\ .
\end{align}
These eigenvectors are the same for any  Higgs sector; 
we verify this  for some particular cases which are  easy to tackle analytically.
By taking  $v_1=v_2$, $b_1=b_3$, $b_2=0$, $M_I=M_R$  and
$M_3=0$, there are two limits $M_I\rightarrow \infty $ 
or  $b_3 \rightarrow \infty $, which do not correspond to 
a realistic potential; however, they  serve us to  check  that
in both cases we obtain the eigenvectors in  Eq.~(\ref{eq:eigenvector1}) and Eq.~(\ref{eq:eigenvector2}). 
It is possible to build Higgs
tensors in the $\bar{3}_L\times 3_L$ and $\bar{3}_R\times 3_R$ representations, these terms give masses to  
$A_{L3}^{U}=-(A_{L3}^I-\sqrt{3}A_{L8}^I)/2$  and $A_{R3}^{U}=-(A_{R3}^I-\sqrt{3}A_{R8}^I)/2$, respectively; 
however, their  contribution to the vector boson mass matrix  do not change the present results.
\section{Alternative left-right models}
\label{sec:alr}
As we already saw in Section~\ref{sec:trinification},  by choosing other $SU(2)$
spin symmetries, it is possible to find alternative models to 
the  left-right Symmetric model which have been studied extensively~\cite{Pati:1974yy,Mohapatra:1974hk,Mohapatra:1974gc,Senjanovic:1975rk,Mohapatra:1979ia}.
The gauge group of the  low energy effective theory is  $G=SU(3)_C\otimes SU(2)_L\otimes SU(2)_X\otimes U(1)_{BLX} $. 
In the literature  the spin symmetry  $SU(2)_{I}$ 
corresponds to  $SU(2)_R$\footnote{We do not use the label $R$ 
for this symmetry because the alternative spin symmetries also are  subgroups of $SU(3)_R$.} 
and $U(1)_{BLI}$ corresponds  to   $U(1)_{B-L}$. 
The resulting model by choosing $X=V$ is  known as 
the alternative left-right model~\cite{Ma:1986we} as we already mentioned in Section~\ref{sec:trinification}. 
The case  $X=U$  is a new model where $U(1)_{BLU}$
corresponds to  the  $E_6$ lephophobic model (modulo a normalization which is  important for the phenomenology). 
In this section we study these models
as  low energy  effective  field theories for $[SU(3)]^3$~\cite{Hetzel:2015bla,Hetzel:2015cca}.  
The neutral current  Lagrangians for these models are
\begin{align}\label{eq:LRLagrangian}
-\mathcal{L}_{NC}=  
 &g_L J_{L3\mu}^{I}A_{L3}^{I\mu}+g_R J_{R3\mu}^{X}A_{R3}^{X\mu}+g_{BL}^X J_{BL\mu}^XA_{BL}^{X\mu}\notag\\
=&g_L J_{L3\mu}^{I}A_{L3}^{I\mu}+g^{\prime}J_{Y\mu} B^\mu+g_2J_{2\mu} Z^{\prime\mu},\hspace{0.05cm}X=I,V\ ,\\
 -\mathcal{L}_{NC}=
 &g_L J_{L3\mu}^{I}A_{L3}^{I\mu}+g_R J_{R8\mu}^{U}A_{R8}^{U\mu}+g_{BL}^U J_{BL\mu}^UA_{BL}^{U\mu}\notag\\
=&g_L J_{L3\mu}^{I}A_{L3}^{I\mu}+g^{\prime}J_{Y\mu} B^\mu+g_2J_{2\mu} Z^{\prime\mu}\ ,\hspace{0.2cm}X=U\ .
\end{align}
For $X=U$,  the weak-$U$-spin operator   $U_{R3}$ does not contribute 
to the charge operator $Q$; so, it is not mandatory to take into account  the corresponding current in the Lagrangian;
however, $U_{R8}$ is necessary in order to reproduce the  electromagnetic charges of the $27$.
For $X=I$ and $X=V$, from the $[SU(3)]^3$ charges we obtain
\begin{align}
Q=&I_{L3}+c_X X_{R3}+X_{BL}\ ,\hspace{1.8cm} X=I,V\ ,\\
Q=&I_{L3}+\frac{2}{\sqrt{3}}(d_U-\frac{1}{2})X_{R8}+X_{BL}\ ,\hspace{0.35cm} X=U\ ,
\end{align}
where  
\begin{align}
  X_{BL} =&  \frac{1}{\sqrt{3}}I_{L8}+\frac{1}{\sqrt{3}}X_{R8}\ ,\hspace{1.5cm}X=I,U,V\ .
\end{align}
The $X_{BLX}$ charges are not $E_6$ normalized 
and as it can be verified in Table~\ref{tab:e6charges},
for $X=I$ these  charges correspond to the 
$(B-L)/2$ ones, \ie $I_{BL}=(B-L)/2$. 
By means of an orthogonal matrix we can rotate from the left-right basis of the NC vector bosons 
to the $(B,Z')$ basis \ie
\begin{align}\label{eq:orthogonalBL}
\begin{pmatrix}
 B_\mu \\
 Z'_\mu 
\end{pmatrix}
=& 
\left(\mathcal{O}_{BL}^{I,V}\right)^T
\begin{pmatrix}
 A_{R3\mu}^{I,V} \\
 A_{LB\mu}^{I,V} 
\end{pmatrix}\ ,
\hspace{0.1cm}\notag\\
\begin{pmatrix}
 B_\mu \\
 Z'_\mu \\
\end{pmatrix}
= &
\left(\mathcal{O}_{BL}^{U}\right)^T
\begin{pmatrix}
 A_{R8\mu}^U \\
 A_{LB\mu}^U \\
\end{pmatrix}\ ,
\end{align}
where the orthogonal matrices  are  
\begin{align*}
\mathcal{O}_{BL}^{I,V}= 
\begin{pmatrix}
\cos \gamma &\quad \sin \gamma \\
\sin \gamma & -\cos \gamma
\end{pmatrix}\ ,
\hspace{0.1cm}
\mathcal{O}_{BL}^{U}= 
\begin{pmatrix}
\cos \delta &\quad \sin \delta \\
\sin \delta & -\cos \delta
\end{pmatrix}
\ .
\end{align*}
By replacing this expression in Eq.~(\ref{eq:LRLagrangian}) we obtain
\begin{align}
g^{\prime}B_\mu J_Y^\mu =& B^\mu\left(g_RJ_{R3\mu}^X \cos \gamma +g_{BL}^X J_{BL\mu}^X \sin \gamma   \right)\ ,   \\
g^{\prime}B_\mu J_Y^\mu =& B^\mu\left(g_RJ_{R8\mu}^X \cos \delta +g_{BL}^X J_{BL\mu}^X \sin \delta   \right)\ ,
\end{align}
For $X=I,V$ and $X=U$ respectively.
 By equating the hypercharge current with 
\begin{align}
J_{Y\mu} =& c_X J_{R3\mu}^X+ J_{BL\mu}^X\ ,\hspace{1.7cm}  X=I,V\ ,   \\
J_{Y\mu} =& \frac{2}{\sqrt{3}}(d_U-\frac{1}{2}) J_{R8\mu}^X+ J_{BL\mu}^X\ ,\ \ X=U\ ,
\end{align}
we get the equations 
\begin{align*}
g_R \cos \gamma =& g^{\prime}c_X\ ,\hspace{1.3cm} g_{BL}^X \sin \gamma =  g^{\prime},\ \ X=I,V\ ,\\ 
g_R \cos \delta =& g^{\prime}\frac{2}{\sqrt{3}}(d_U-\frac{1}{2})\ ,\hspace{0.2cm} g_{BL}^X \sin \delta =  g^{\prime}, \ \ X=U\ .
\end{align*}
From these equations we get
\begin{align*}
 \cos \gamma =& c_X\frac{g^\prime}{g_R}\ ,
 \hspace{0.6cm}\frac{1}{g^{\prime\,2}}=\frac{1}{\left(g_{BL}^{X}\right)^2}+\frac{c_X^2}{g_R^2}\ ,\hspace{0.2cm}X=I,V \ . \\
 \cos \delta =& -\sqrt{3}\frac{g^\prime}{g_R}\ ,
 \hspace{0.2cm}\frac{1}{g^{\prime\,2}}=\frac{1}{\left(g_{BL}^{X}\right)^2}+\frac{3}{g_R^2}\ ,\hspace{0.2cm}X=U\ .
\end{align*}
Because  $c_X=1$ for  $X=I,V$ the right gauge coupling must satisfy the inequality  $g_R>g^{\prime}=0.357$,  
which is met in $[SU(3)]^3$.
For $X=U$  the last equation implies $g_R > \sqrt{3}g^{\prime}=0.619$, which automatically 
excludes the $[SU(3)]^3$ value of the right gauge coupling, \ie  $g_R=0.435$; 
however, the typical left-right gauge coupling $g_L=g_R=0.652$ is still possible.
From equations ~(\ref{eq:LRLagrangian}) and ~(\ref{eq:orthogonalBL}) it is possible 
to get expressions for the neutral current associated with the $Z'$
\begin{align}\label{eq:lrcurrent}
 g_2 J_{2\mu}=& -g_{BL}^X J_{BL\mu}^X \cos \gamma + g_R J_{R3\mu}^X \sin \gamma\notag\\     
             =& g_L \tan_W \left(\alpha_X J_{R3\mu}^X- \frac{c_XJ_{BL\mu}^X}{\alpha_X}\right),\ X=I,V \ ,\notag \\
 g_2 J_{2\mu}=& -g_{BL}^X J_{BL\mu}^X \cos \delta + g_R J_{R8\mu}^X \sin \delta\notag\\     
             =& g_L \tan_W \left(\alpha_U J_{R8\mu}^X+\frac{\sqrt{3}J_{BL\mu}^X}{\alpha_U}\right),\ X=U \ ,
\end{align}
where
\begin{align}\label{eq:alphax}
\alpha_X=&\sqrt{\left(\frac{g_R}{g_L}\right)^2\cot^2\theta_W-c^2_X}
        \ \underrightarrow{[SU(3)]^3}\notag\\  
        &\frac{1}{\sqrt{4\cos^2\theta_W-1}}\ ,\hspace{1.5cm}\ X=I,V\ ,         \ \notag\\
\alpha_U=&\sqrt{\left(\frac{g_R}{g_L}\right)^2\cot^2\theta_W-3}\ ,\hspace{0.7cm} X=U\ .        
\end{align}
From these expressions we can obtain the explicit expressions for the 
vector and axial charges.
For $X=I$ and $g_R= g_L$ these charges correspond to those 
of the left-right model reported in reference~\cite{Langacker:2008yv}. 
In the present work $g_R$ is determined by the $[SU(3)]^3$ symmetry, thus, 
by replacing $g_R$  from Eq.~(\ref{eq:gr}) in the expression above we obtain~(for $X=I,V$) 
the r.h.s expression of Eq.~(\ref{eq:alphax}). For $\sin^2 \theta_W = 3/8$ we recover 
the unification matching condition $g_L=g_R$ for any $X$. However, in  the present 
work we make use of the  $\overline{\text{MS}}$ value for the weak mixing angle,  $\sin\theta_W= 0.231$,
as we will explain below. 
From these relations and from Table~\ref{tabla:1}  we can obtain the  explicit expressions for the vector and axial charges
for the $Z'$ gauge boson, corresponding to the  $g_2J_{2\mu}$ current. For $X=I,V,U$ these charges are shown in the Tables~\ref{tab:lrmodel},
\ref{tab:alrmodel} and \ref{tab:notlrmodel}, respectively.
The collider and EW constraints are shown in Table~\ref{tab:limits} and Figures~\ref{Contours2} and \ref{Contours3}.
A detailed analysis of these constraints will be  presented  in the next  Section. 

\section{Electroweak and collider constraints}
\label{sec:ewconstraints}
\begin{table} 
\centering
\begin{tabular}{|l|r|r|c|c|c|} 
\hline
$Z'$           & \multicolumn{2}{c|}{$M_{Z'}$ [GeV]} & \multicolumn{3}{c|}{$\sin\theta_{ZZ'}$}   \\ \hline
                    &  LHC  &  EW   & $\sin\theta_{ZZ'}$ & $\sin\theta_{ZZ'}^{\rm min}$ & $\sin\theta_{ZZ'}^{\rm max}$  \\ 
\hline
$Z_{331G}$             & 2,925 &  958   & $-0.00007$  & $-0.0012 $ & 0.0009     \\ \hline
$Z_I^{\text{Tri}}$     & 2,492 & 1,134 &  $ 0.0003$ & $-0.0006$ & 0.0013  \\ 
$Z_I$                  & 2,525 & 1,204 &  $ 0.0003$ & $-0.0005$ & 0.0012  \\ \hline
$Z_{LR}^{\text{Tri}}$  & 2,693 & 1,182 &  $-0.0004$ & $-0.0015$ & 0.0006  \\ 
$Z_{LR}$               & 2,682 &  998  &  $-0.0004$ & $-0.0013$ & 0.0006  \\ \hline
$Z_{LRU}$              & 2.588 &  935  &  $-0.00001$& $-0.0011$  & 0.0008  \\  \hline
$Z_{ALR}^{\text{Tri}}$ & 2,532 &  447  &  $-0.0004$ & $-0.0014$ & 0.0007  \\ \hline

\end{tabular}
\caption{95\% C.L. lower mass limits on extra $Z'$ bosons for various models 
from EW precision data and constraints on $\sin\theta_{ZZ'}$. 
For comparison, we show  in the second column 
the 95\% LHC constraints~\cite{Aad:2014cka} which have been calculated according to Ref.~\cite{Salazar:2015gxa}. In the 
following columns we give, respectively, 
the central value and the 95\% C.L. lower 
and upper limits for $\sin \theta_{ZZ'}$.}
\label{tab:limits}
\end{table}

We analyze  the previously considered neutral gauge bosons and impose limits on 
the $Z\text{-}Z'$ mixing angle,  $\theta_{Z-Z'}$,  and on the masses of the neutral $Z'$ bosons, $M_{Z'}$.
In order to obtain the EW Precision Data~(EWPD) constraints,  we make use 
of the special purpose FORTRAN package, 
GAPP~(Global Analysis of Particle Properties)~\cite{Erler:1999ug}.
Details of the  analysis can be found  in Ref.~\cite{Erler:2009jh,Erler:2009ut,Erler:2010uy,Erler:2011iw}
\footnote{An update of Ref.~\cite{Erler:2009jh} will be presented soon.}.\\

In the third column of  Table~\ref{tab:limits}    the EW constraints are shown. 
The quantum numbers  of the  model $Z_{331G}$ correspond to those of $U(1)_{21\overline{I}}$
 in Table~\ref{tab:e6charges}.
We do not put the  superscript $\text{Tri}$ on the 3-3-1 model because the charges and the coupling 
   strength of this model are the same as 
   the very well known universal 3-3-1 model~\cite{Singer:1980sw,Sanchez:2001ua,Erler:2011ud}, or the so called  $G$ model in 
references~\cite{Salazar:2015gxa,Ponce:2001jn}.
The vector and axial charges for this model are shown in Table~\ref{tab:331g}.

The  quantum numbers of   $Z^{\text{Tri}}_{I}$ correspond to those of $U_{I}$ in Table~\ref{tab:e6charges}.  
This model is known as the inert model which does not couple to up-type quarks~\cite{Robinett:1982tq},
and corresponds to the second  neutral vector boson or $Z^{\prime\prime}$  in the $[SU(3)]^3$ group.
From Eq.~(\ref{eq:currents}) for $X=U$ we can see that the coupling strength of $Z_I^{\text{Tri}}$
is  $g_2=g_R=0.435$. To get this number in  Eq.~(\ref{eq:gr}) we use  for the  weak mixing angle   the value  $\sin \theta_W = 0.231$,
which corresponds to the $\overline{\text{MS}}$ renormalization scheme at the $Z$-pole scale. 
This value is different of the traditional $E_6$  coupling strength,
$g_2= \sqrt{\frac{5}{3}}g_L\tan \theta_W= 0.4615$. 
The   constraints by using the $E_6$ coupling strength correspond to those of  $Z_I$ in~Table~\ref{tab:limits}. 
The inequality of the couplings is reflected 
in the EW and LHC constraints. The axial and vector couplings of this model are shown in Table~\ref{tab:inertmodel}.

In Table~\ref{tab:limits} we also distinguish between 
$Z_{LR}$, which assume the equality between the left and right gauge couplings, \ie  $g_R=g_L=0.652$, 
 and $Z_{LR}^{\text{Tri}}$  for which the 
right coupling strength is dictated by $[SU(3)]^3$, \ie $g_R=0.435$.  
This inequality  between the left and right couplings 
makes the chiral charges different which is the reason of 
the disparity in the constraints in Table~\ref{tab:limits}.

We observe that for the alternative left-right model~(ALR) the EW constraints
are weak compared to other typical $E_6$ models in the literature 
(except the   $Z_\psi$ which  only has  axial couplings to the SM particles   and the leptophobic model $Z_{\not{L}}$),
which  is a well known feature of this model~\cite{Erler:1999nx}. 
  
As we already saw in Section~\ref{sec:alr}, there is another alternative model for $I_{BL}=(B-L)/2$,  
the $U_{BL}$  which, to the best of our knowledge, has not been studied  before. 
This model is $U$-spin symmetric [\ie  $SU(2)_{U}$ ], and it has as 
the main feature that it is leptophobic in the limit $g_R\rightarrow 0.619^+$.
In the aforementioned limit   $\alpha_U\rightarrow 0$
and the lepton couplings are proportional to $\alpha_U$; 
 however, because in this limit the quarks couplings go as $\sim 1/\alpha_U$, 
 in many observables  these effects compensate each other in such a way 
 that  the  EW constraints are not trivial  for this  model.   
\begin{figure*}[h]
\centering 
\begin{tabular}{cc}
\includegraphics[scale=0.30]{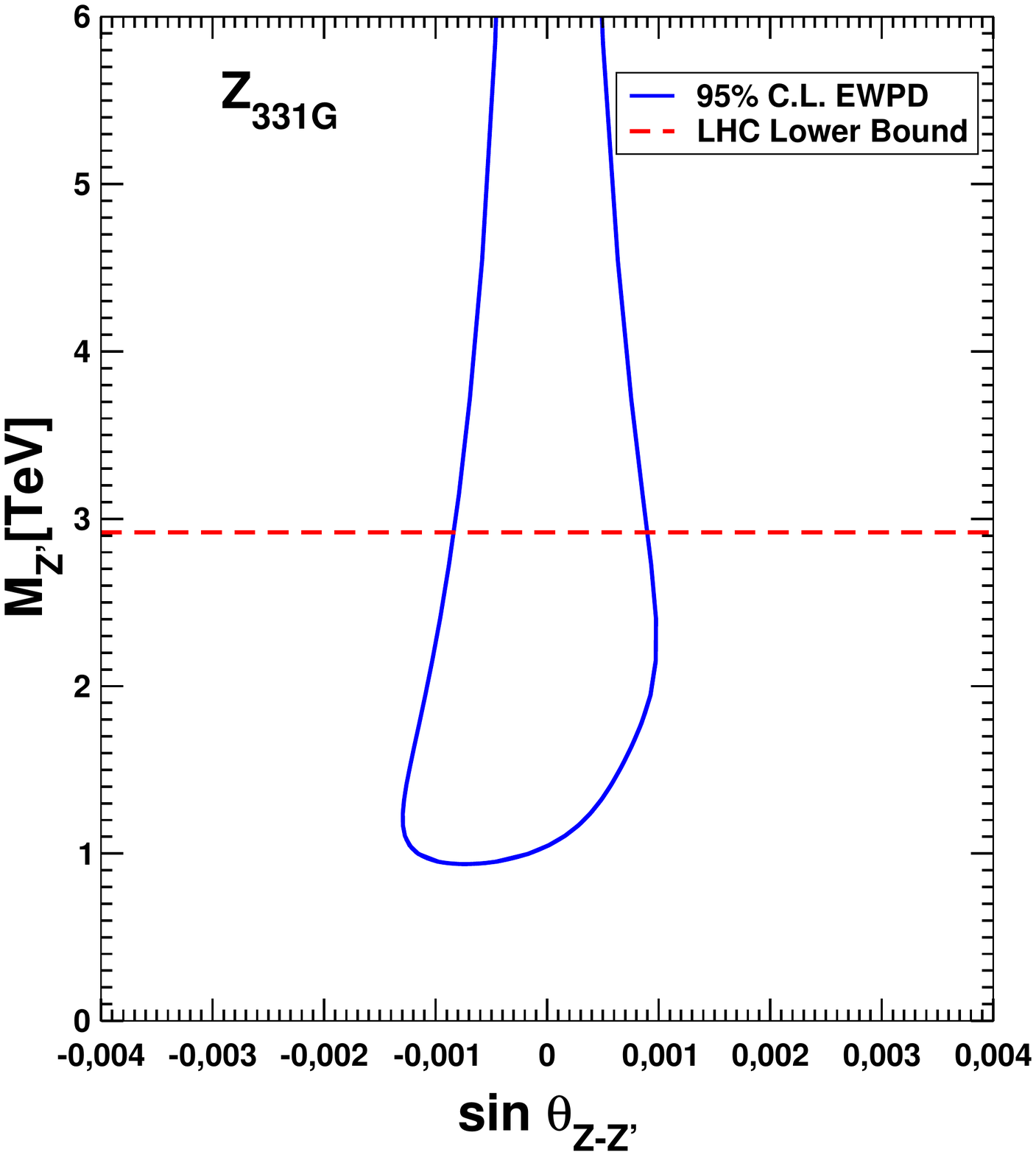} & \includegraphics[scale=0.30]{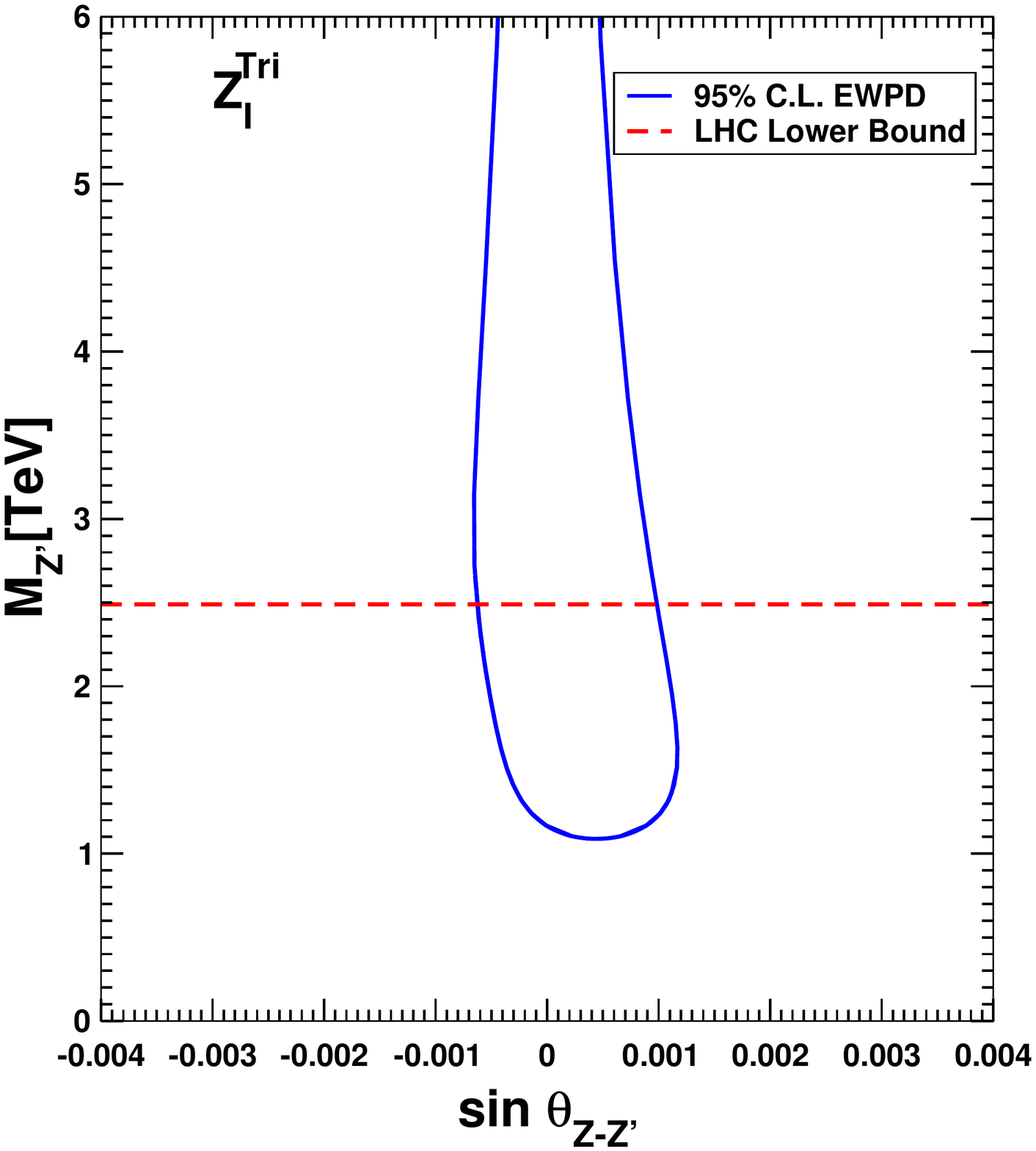} \vspace{-12pt} 
\end{tabular}
\caption{The continuous blue line represents the 90\% C.L.  exclusion  contour in $M_{Z'}$ vs. $\sin \theta_{ZZ'}$ for
the universal 3-3-1 model which has charges and coupling strength according to Eq.~(\ref{eq:currents})
with $X=I$. 
The axial and vector charges for this model are shown in Table~\ref{tab:331g}. The inert model  $Z_{I}^{\text{Tri}}$ has 
the same charges as the $E_6$ motivated  $Z_I$, but with  the coupling strength dictated by $[SU(3)]^3$ according to 
 Eq.~(\ref{eq:currents}) for  $X=U$, \ie  $g_3=g_R=435$.The axial and vector charges for this model are shown in Table~\ref{tab:inertmodel}.
The corresponding plot for the $E_6$ motivated  $Z_I$ is shown in Ref.~\cite{Erler:2009jh}.
The red dashed line is   the 95\% C.L. lower mass limit obtained from ATLAS data~\cite{Aad:2014cka}. }
\label{Contours}
\end{figure*}

\begin{figure*}
\centering 
\begin{tabular}{cc}
\includegraphics[scale=0.30]{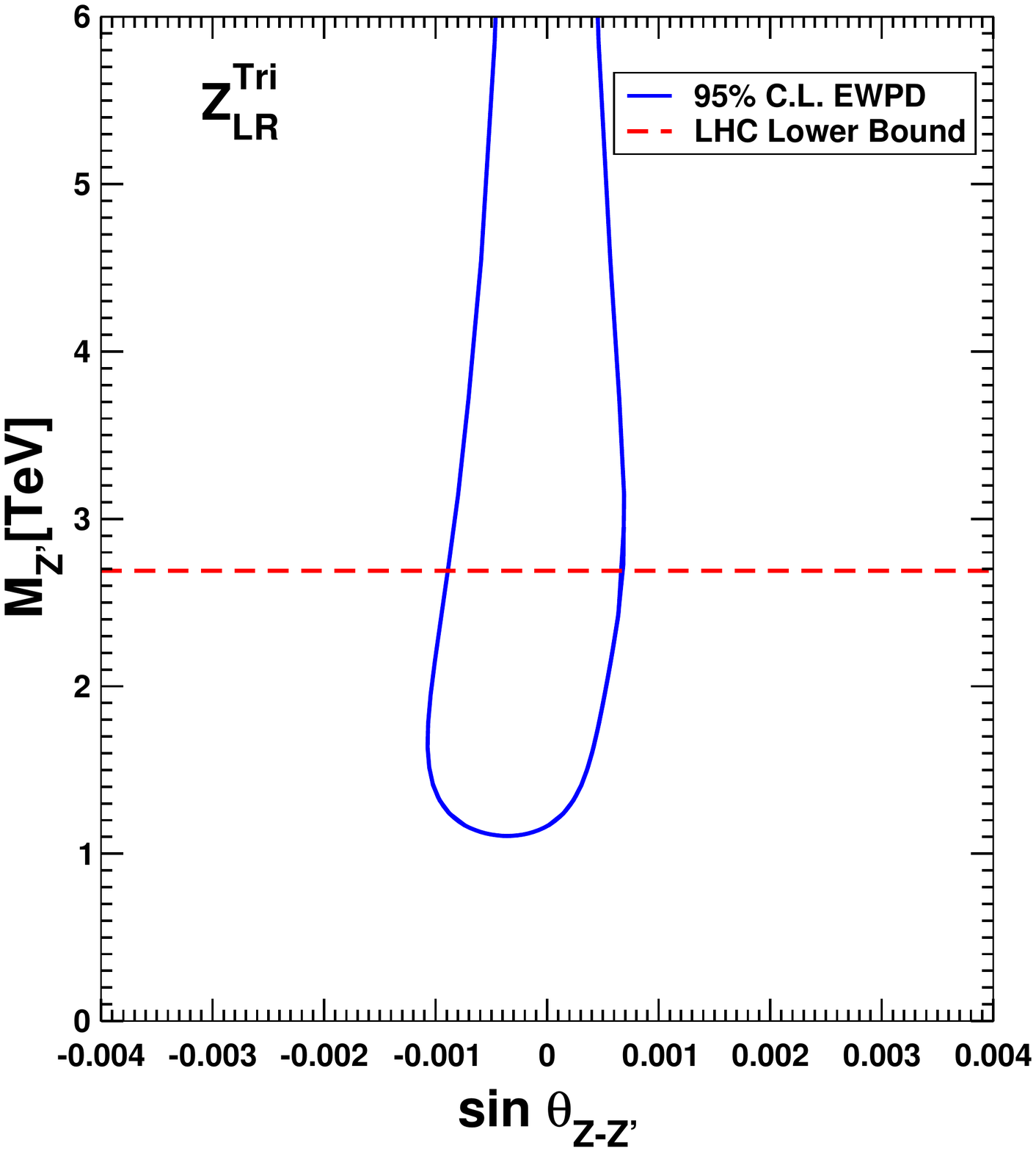} & \includegraphics[scale=0.30]{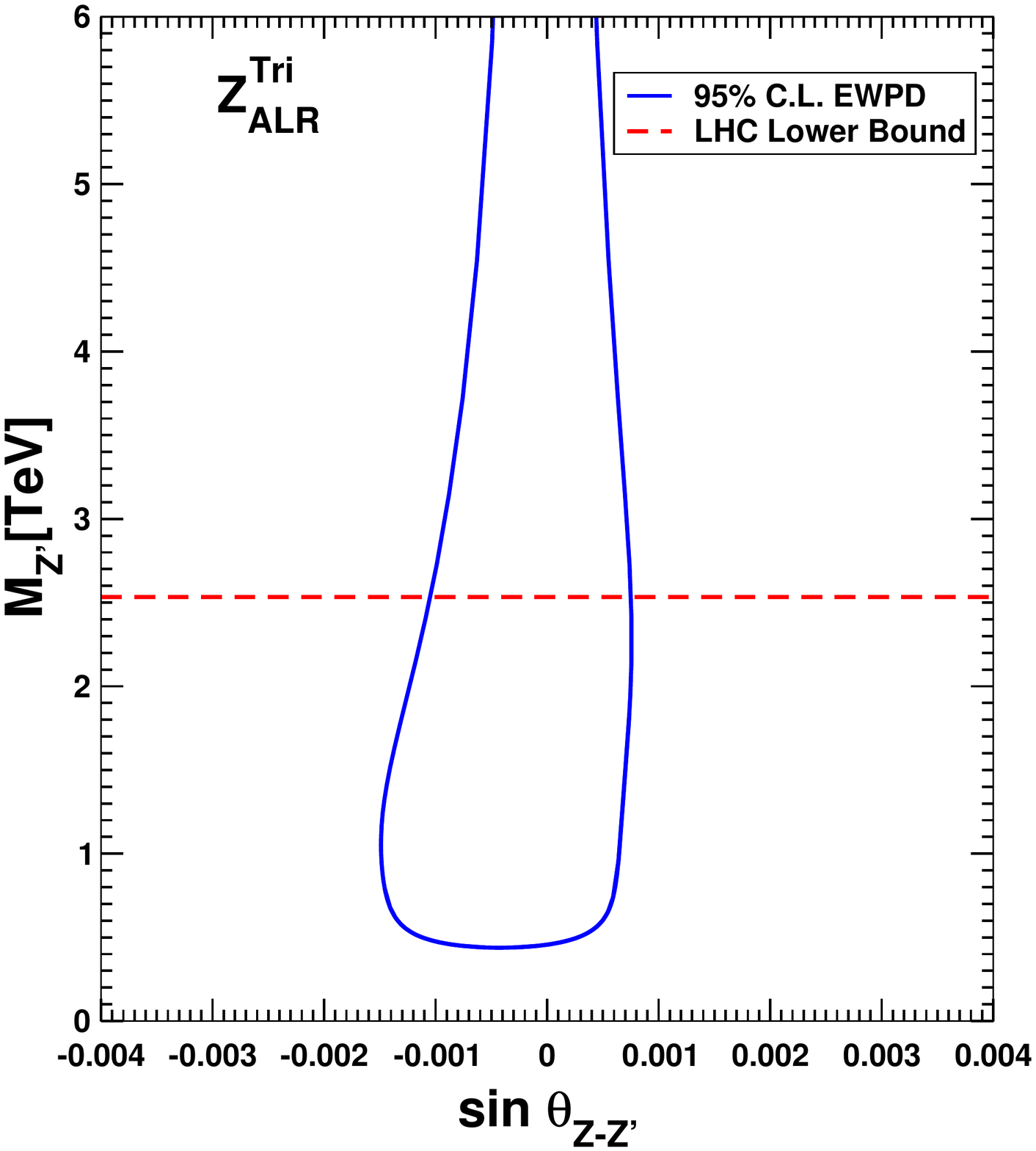} \\
\end{tabular}
\caption{
The continuous blue line represents the 90\% C.L.  exclusion  contour in $M_{Z'}$ vs. $\sin \theta_{ZZ'}$ for
 the left-right symmetric model $Z_{LR}^{\text{Tr}}$ and the Alternative left-right Model $Z_{ALR}^{\text{Tr}}$ 
 with the right coupling strength dictated by $[SU(3)]^3$, \ie  $g_R=0.435$ for 
 $\sin \theta_W= 0.231$~(see Eq.~(\ref{eq:lrcurrent}) for $X=I$ and $X=V$, respectively).
 The axial and vector charges for the left-right and the ALR model are shown in Table~\ref{tab:lrmodel} and Table~\ref{tab:alrmodel}.
The corresponding plot for the left-right symmetric model  with $g_R=g_L=0.652$ is shown in Ref.~\cite{Erler:2009jh}.
The red dashed line is   the 95\% C.L. lower mass limit obtained from ATLAS data~\cite{Aad:2014cka}. }
\label{Contours2}
\end{figure*}

\begin{figure*}
\centering 
\begin{tabular}{c}
\includegraphics[scale=0.38]{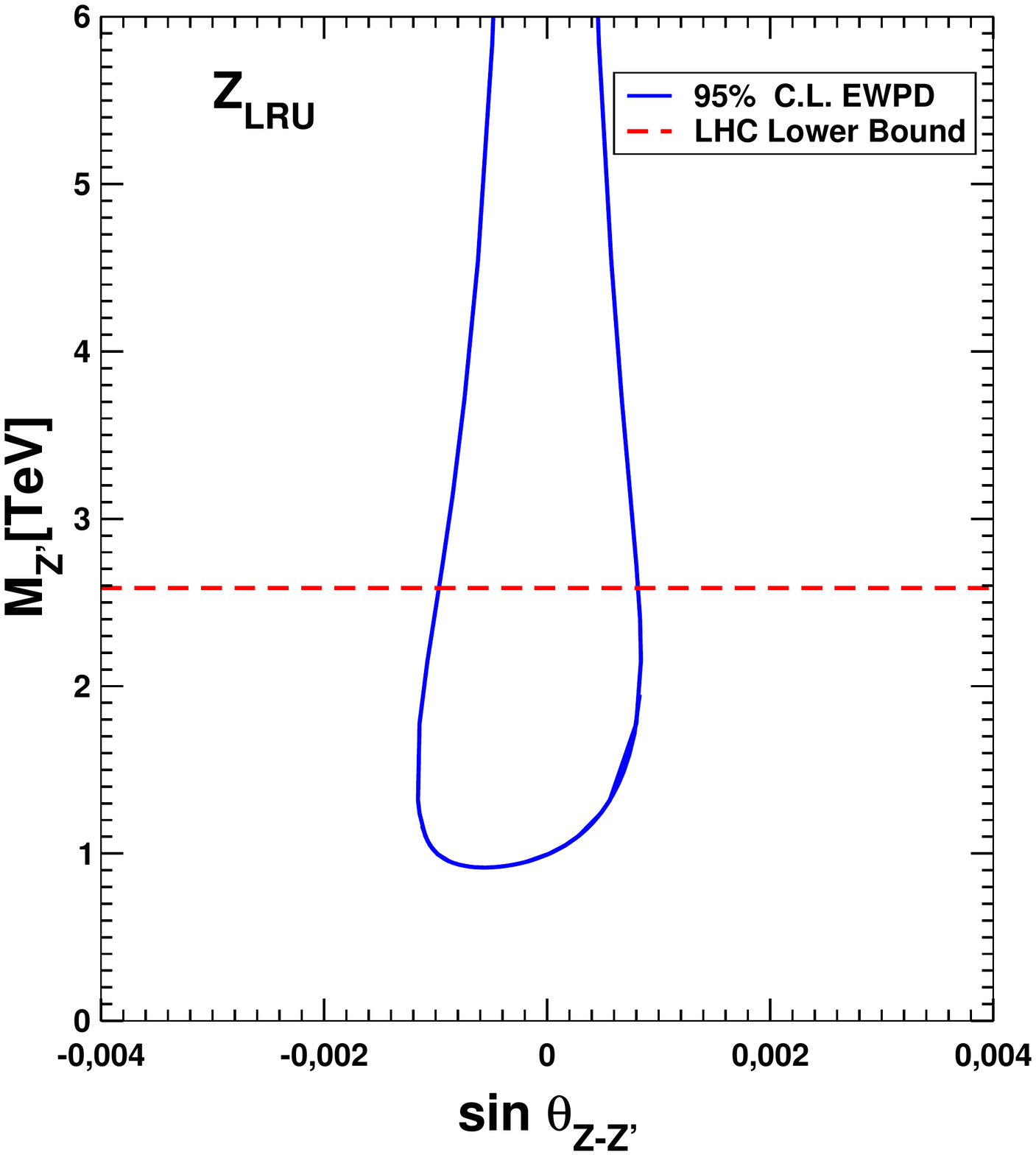} \vspace{-12pt} 
\end{tabular}
\caption{
The continuous blue line represents the 90\% C.L.  exclusion  contour in $M_{Z'}$ vs. $\sin \theta_{ZZ'}$ for
 the LRU model $Z_{LRU}^{\text{Tr}}$  with    $g_R=g_L=0.652$ for 
 $\sin \theta_W= 0.231$~(See Eq.~(\ref{eq:lrcurrent}) for  $X=U$ ).
 The axial and vector charges for the left-right and the LRU model are shown in Table~\ref{tab:lrmodel} and Table~\ref{tab:alrmodel}.
The red dashed line is   the 95\% C.L. lower mass limit obtained from ATLAS data~\cite{Aad:2014cka}.}
\label{Contours3}
\end{figure*}

In Figures~\ref{Contours} and \ref{Contours2} the 90\% exclusion contours  for the universal  3-3-1 model $Z_{331G}$,
 its corresponding $Z^{\prime\prime}= Z_{I}^{\text{Tri}}$ in the $[SU(3)]^3$ model, in   
 the left-right symmetric model  $Z_{LR}^{\text{Tri}}$, and in its alternative version 
the $Z_{ALR}^{\text{Tri}}$    are shown.   
The plots for   $Z_{LR}^{\text{Tri}}$ and
the inert model $Z_{I}^{\text{Tri}}$ are comparable  with  $Z_{LR}$ and 
$Z_{I}$ in reference~\cite{Erler:2009jh}. 
Because $g_R> 0.619$ as we already saw in Section~\ref{sec:alr}, 
it  is not possible to have  the $Z_{LRU}$ coming from  a 
low energy  $[SU(3)]^3$ effective model; however, by choosing  $g_R=g_L=0.652$ 
this model is feasible.  The corresponding EW and 
LHC constraints are shown in Table~\ref{tab:limits} and Figure~\ref{Contours3}.

In Ref~\cite{Aad:2014cka} the ATLAS detector data  on dilepton production 
was used to search for high-mass resonances decaying to dielectron or dimuon final states. 
The experiment analyze
proton-proton collisions at a center of mass energy of 8 TeV and an integrated luminosity of
20.3~$fb^{-1}$ in the dielectron channel, and 20.5~$fb^{-1}$
in the dimuon channel. From this data they
report 95\% CL upper limits on the total cross-section of 
$Z^0$ decaying to dilepton final states. 
From these results, and following our earlier analysis~\cite{Salazar:2015gxa}, 
we obtain the 95\% C.L. lower mass limits  for all the models mentioned above.
These limits  are shown in  the  second column in
Table~\ref{tab:limits}  and they correspond to the red dashed line 
in Figures~\ref{Contours}  and \ref{Contours2}.

\section{Conclusions}
\label{sec:conclusions}
In this work we analyzed all the possible 
embeddings of the 3-3-1 and 3-2-2-1 models present  in the $[SU(3)]^3$ gauge group.
 By considering the   weak-$U$-spin and weak-$V$-spin symmetries in $SU(3)_R$ besides the usual 
weak-$I$-spin  symmetry [best known as $SU(2)_R$] we found two flipped versions of the 
3-3-1 model, with the particularity that  the $Z'$ axial and vector  charges 
are identical for the three  spin symmetries;   hence, they are not a new source of  phenomenological results. 
In  Appendix~\ref{appendixa}, we showed  that the reason behind  these results is that, 
just  for these models,  the corresponding neutral current Lagrangians 
are related each other by unitary transformations.
 For the  left-right symmetric model we also found two flipped versions one of them
 not reported in the literature as far as we know. 
This new  model is denoted as $Z_{RLU}$ and it corresponds to  a second alternative model of the 
left-right  model  $Z_{LR}$ (the first alternative model  is $Z_{ALR}$ which is  well known in the literature~\cite{Ma:1986we}).
 In several respects the $Z_{LRU}$ model is different of $Z_{LR}$ and $Z_{ALR}$; 
for example, it is not viable as a low energy effective theory, unless we make it 
left-right symmetric, which is a typical  assumption of the $Z_{LR}$ and $Z_{ALR}$ models. 
This model  has as  the main feature that it is leptophobic in the limit $g_R\rightarrow 0.619^+$.
In the aforementioned limit   $\alpha_U\rightarrow 0$
and the lepton couplings are proportional to $\alpha_U$;  however, because in this limit the quarks couplings go as $\sim 1/\alpha_U$, 
 in many observables  these effects compensate each other in such a way 
 that  the  EW constraints are not trivial. 
 
 We also calculated the eigenstates of the   $[SU(3)]^3$ Higgs potential and,
by considering different cases, 
it was shown  that these eigenstates are independent of the Higgs sector. 
It was also shown that the null space of the $[SU(3)]^3$ vector boson mass matrix  
corresponds to the photon. As a generalization of these results,  
we gave the explicit form of  the null vector of  the EW
vector boson mass matrix  for  an arbitrary Higgs tensor and an arbitrary  gauge group.  

By using  the LHC experimental results  and EW precision data, 
new limits on the $Z'$ mass $M_{Z'}$  and the mixing angle $\theta_{Z-Z'}$ are 
imposed. From this analysis we found lower limits 
on $M_{Z'}$ of the order of $2.5$~TeV, while the mixing angle 
was found to be constrained to values of the  order of $10^{-3}$~radians.\\

The scope of the present work is not limited to the $[SU(3)]^3$ group.
In reference~\cite{Rojas:2015tqa} the full  set of alternatives breakings in $E_6$ was shown, the next step is to extend 
our analysis  to $E_6$, which has as subgroups  
the most promising and best-known electroweak extensions of the standard model.

\section*{Acknowledgments} 
We thank financial support from 
\enquote{Patrimonio Autónomo Fondo Nacional de Financiamiento para la Ciencia, la Tecnología
y la Innovación, Francisco José de Caldas}, 
and \enquote{Sostenibilidad-UDEA 2014-2015}.
 R. H. B. thanks to \enquote{Centro de Investigaciones del ITM}.
\appendix
\section{The weak-I, weak-U and weak-V spin symmetries } 
\label{sec:su3sctruture}
\begin{table}
\begin{center}
\bgroup
\def\arraystretch{1.5}%
\begin{tabular}{|cccccc|cccccc|}
\hline
       \multicolumn{6}{|c|}{$U\lambda_i U^{\dagger}=\lambda^U_i $ }&       \multicolumn{6}{|c|}{$V\lambda_i V^{\dagger}=\lambda^V_i $ }\\
\hline        
       $\lambda_1^{U}$&$\lambda_2^{U}$&$\lambda_4^{U}$ &$\lambda_5^{U}$& $\lambda_6^{U}$ & $\lambda_7^{U}$&  
       $\lambda_1^{V}$&$\lambda_2^{V}$&$\lambda_4^{V}$ &$\lambda_5^{V}$& $\lambda_6^{V}$ & $\lambda_7^{V}$\\
\hline
       $\lambda_6 $   &$\lambda_7 $   &$\lambda_1 $    &$-\lambda_2 $  & $\lambda_4 $    & $-\lambda_5 $ &
       $\lambda_4 $   &$-\lambda_5$   &$\lambda_6 $    &$-\lambda_7 $  & $\lambda_1 $    & $ \lambda_2 $\\
 \hline 
\end{tabular}
\caption{The $SU(3)$ algebra is invariant under a unitary transformation. 
By requiring  that  $\lambda_3$  and $\lambda_8$ be mapped to  diagonal matrices 
there are two possible choices, U and $V=U^{\dagger} $.  Additionally, these 
matrices satisfy  $U^2 = U^{\dagger}$, and $V^2 = V^{\dagger}$; from these relations 
and unitarity we obtain $U^3=V^3= 1$. The latter identity allows us to verify  
the Table entries.}
\label{tab:lambdauv}
\egroup
\end{center}
\end{table}
The $SU(3)$ algebra is invariant under any unitary transformation. \ie
\begin{align*}
[\lambda_a/2,\lambda_b/2]= if_{abc}\lambda_c/2 \rightarrow
[\lambda_a^{\prime}/2,\lambda_b^{\prime}/2]= if_{abc}\lambda_c^{\prime} /2  \ , 
\end{align*}
where  $\lambda_a^{\prime} = U\lambda_a U^{\dagger}$.  By requiring  
that  $\lambda_3$  and $\lambda_8$ be mapped to  diagonal matrices 
a   form of the unitary matrices is~(there are several ways to choose U and V) 
\begin{align*}
 U= 
 \begin{pmatrix}
 0 & 0 & 1\\ 
 1 & 0 & 0\\
 0 & 1 & 0\\
 \end{pmatrix}\ ,
 \hspace{1cm}
  V= 
 \begin{pmatrix}
 0 & 1 & 0\\ 
 0 & 0 & 1\\
 1 & 0 & 0\\
 \end{pmatrix}
 \ .
\end{align*}
Additionally, these 
matrices satisfy  $U^2 = U^{\dagger}$, and $V^2 = V^{\dagger}$; from these relations
and unitarity we obtain $U^3=V^3= 1$. 
The operators corresponding to $\lambda_3$ and $\lambda_8$ Gell-Mann   generators for  the weak-$I$-spin~(or Isospin),   weak-$U$-spin and weak-$V$-spin are
\begin{align*} 
\lambda_3^{I}=& \lambda_3 = 
\begin{pmatrix}
  1 & 0 & 0 \\
  0 & -1 & 0 \\ 
  0 & 0 &  0 \\ 
\end{pmatrix}\ ,
\notag\\
\lambda_3^{U}=& U \lambda_3 U^{\dagger}=
-\frac{1}{2}\left(\lambda_3-\sqrt{3}\lambda_8\right)
=
\begin{pmatrix}
  0 & 0 & 0 \\
  0 & 1 & 0 \\ 
  0 & 0 & -1 \\ 
\end{pmatrix}\ ,
\notag\\
\lambda_3^{V}=&V \lambda_3 V^{\dagger}=
-\frac{1}{2}\left(\lambda_3+\sqrt{3}\lambda_8   \right)
=
\begin{pmatrix}
 -1 & 0 & 0 \\
  0 & 0 & 0 \\ 
  0 & 0 & 1 \\ 
\end{pmatrix}\ ,
\notag\\ 
\end{align*}
and 
\begin{align*} 
\lambda_8^I =& \lambda_8 
=\frac{1}{\sqrt{3}}
\begin{pmatrix}
  1 & 0 & 0 \\
  0 & 1 & 0 \\ 
  0 & 0 & -2 \\ 
\end{pmatrix}
\ ,\notag\\
\lambda_8^{U}=& U \lambda_8 U^{\dagger}
=-\frac{1}{2}\left(\lambda_8+\sqrt{3}\lambda_3   \right)
=\frac{1}{\sqrt{3}}
\begin{pmatrix}
  -2 & 0 & 0 \\
  0 & 1 & 0 \\ 
  0 & 0 & 1 \\ 
\end{pmatrix}
\ ,\notag\\
\lambda_8^{V}=& V \lambda_8 V^{\dagger}
=-\frac{1}{2}\left(\lambda_8-\sqrt{3}\lambda_3   \right)
=\frac{1}{\sqrt{3}}
\begin{pmatrix}
  1 & 0 & 0 \\
  0 & -2 & 0 \\ 
  0 & 0 & 1 \\ 
\end{pmatrix}
\ .\notag\\ 
\end{align*}
Upon the unitary transformations $U$ and $V$,  
the  $SU(3)$ Gell-Mann  matrices $\lambda_i$ 
are mapped to  $\lambda_i^{U}$ and $\lambda_i^{V}$ as it is shown in Table~\ref{tab:lambdauv}.
These alternative representations for the $SU(3)$ algebra are relevant only  for $SU(3)_R$ in the $[SU(3)]^3$ group. 
The representation of the Gell-Mann matrices in   $SU(3)_L$ is fixed by the phenomenology of the SM.\\

\section{U and V in the adjoint representation}
\label{appendixa}
The eight gauge bosons associated with the $SU(3)_R$ are written by convenience as
\begin{align*}
\frac{1}{2}A_{a\mu}^{I}\lambda_{a}^{I}=&
\frac{1}{2}A_{b\mu}^{I} \overline{U}_{\spa\ a}^{Tb} U_a^{\spa c}\lambda_{c}^{I}\\  
=& \frac{1}{2}\left(U_a^{\spa b}A_{b\mu}^{I}\right) \left( U_a^{\spa c}\lambda_{c}^{I}\right)\equiv
\frac{1}{2}A^{U}_{a\mu}  \lambda^{U}_a\ ,
\end{align*}
where the bar in $\bar{U}$  stands for complex conjugation.
Because $U$ is real $U_{a}^{\spa b}=\overline{U}_{a}^{\spa b}$.
In the adjoint representation  $U$ is a $8\times 8$ matrix; 
however, it is reducible to a couple of $3\times3$ matrices  and one  $2\times 2$  matrix.
 The three-dimensional matrices mix the generators associated with the charged bosons
while  the two-dimensional one  mix  the diagonal generators associated with the neutral ones
\begin{align*}
&
\begin{pmatrix}
\lambda_1 \\
\lambda_4 \\
\lambda_6\\
\end{pmatrix}
\xrightarrow{U}
\begin{pmatrix}
\lambda_1^U \\
\lambda_4^U \\
\lambda_6^U\\
\end{pmatrix}
=
\begin{pmatrix}
\lambda_6 \\
\lambda_1 \\
\lambda_4\\
\end{pmatrix}
\ , 
\hspace{1cm}\\
&
\begin{pmatrix}
\lambda_2 \\
\lambda_5 \\
\lambda_7\\
\end{pmatrix}\ 
\xrightarrow{U}
\begin{pmatrix}
\lambda_2^U \\
\lambda_5^U \\
\lambda_7^U\\
\end{pmatrix}
=
\begin{pmatrix}
\lambda_7 \\
-\lambda_2 \\
-\lambda_5\\
\end{pmatrix}\ .
\end{align*}
The  diagonal generators are mapped to 
\begin{align*}
\begin{pmatrix}
\lambda_3 \\
\lambda_8 \\
\end{pmatrix}
\xrightarrow{U}
 \begin{pmatrix}
\lambda_3^U \\
\lambda_8^U \\
\end{pmatrix}
=-\frac{1}{2}
\begin{pmatrix}
\lambda_3-\sqrt{3}\lambda_8\\
\lambda_8+\sqrt{3}\lambda_3\\
\end{pmatrix}
\ .
\end{align*}
We want to make use of this symmetry to rewrite the neutral current
\begin{align}\label{eq:ja}
J_{a\mu}^{I}A_a^{I\mu}=
J_{b\mu}^{I}\overline{U}_{\spa\ a}^{T b} U_a^{\spa c}A_c^{I\mu}=
J_{b\mu}^{I}\overline{V}^{T b}_{\spa\ a} V_a^{\spa c}A_{c\mu}^I\ ,  
\end{align}
by defining  
\begin{align*}
  A_a^{U\mu}   \equiv&  U_a^{\spa c}A_c^{I\mu}\ ,          \hspace{1.9cm}A_a^{V\mu}   \equiv V_a^{\spa c}A_c^{I\mu} ,\notag\\
              J_{a\mu}^{U}\equiv& J_{b\mu}^{I}\overline{U}_{\spa\ a}^{T b}=U_{ a}^{\spa b}J_{b\mu}^{I}
\ ,\hspace{0.5cm}J_{a\mu}^{A}\equiv  J_{b\mu}^{I}\overline{V}_{\spa\ a}^{T b}=V_{ a}^{\spa b}J_{b\mu}^{I}\ , 
\end{align*}
where we take into account that $U$ is a real matrix. By replacing these results in  (\ref{eq:ja}) we obtain
\begin{align}
J_{a\mu}^{I}A_a^{I\mu}=
J_{a\mu}^{U}A_a^{U\mu}=
J_{a\mu}^{V}A_a^{V\mu}\ .
\end{align}
With these expressions it is possible to build the Lagrangian term  $-\mathcal{L}_I=g_R J_{R8\mu}^{X}A_{R8}^{X\mu}$. 
It is important to stress that for the 3-3-1-1 models in  $[SU(3)]^3$  the neutral current Lagrangians of the  alternative models 
are related each other  by  a  unitary transformation; however, in general,  that is not true 
for alternative models. 

\section{The null space of the vector Boson mass matrix for an arbitrary Higgs representation and gauge group.}
\label{appendixb}
In this section we will show that 
for any Higgs potential  there  is  a null 
vector for the  mass matrix $M^{ab}$ of the neutral gauge vector bosons. 
The explicit form of the vector is\footnote{Modulo a normalization.} 
$A^a_{\mu}= \frac{c^a}{g^a} A(x)_{\mu}$,
where the $c^a$ are the coefficients of 
the group generators in the charge operator, \ie
 $Q= c^aT^a$,  the $g^a$ is the coupling strength 
 associated with the $A^a_{\mu}$ vector field and $A(x)_{\mu}$ 
  must be identified with the photon field. 
 For a simple group all the $g^a$ are identical;
 however, they may be  different  for semisimple  groups.
\subsection{Rank 1 tensors} 
 For a rank 1 tensor we can obtain  the vector mass matrix
  from the Higgs covariant derivative
 \begin{align}\label{eq:theorem1}
 \mathcal{L}_{\text{K}}=\text{Tr}\left( (D_\mu \phi^i)^\dagger D_\mu \phi^i\right)    \lvert_{\phi^i=v^i } 
 =\frac{1}{2} A^a_{\mu} M^{ab}A^{b \mu}\ ,
 \end{align}
 where $v^i$ are the components of the vacuum expectation value  vector. 
 This vector  satisfies $Q.v=0$ since the charge operator must annihilate the vacuum. 
 By taking the components of the vector boson as  $A^a_{\mu}= \frac{c^a}{g^a} A(x)_{\mu}$,
 the covariant  derivative becomes zero
 \begin{align*}
  &D_\mu \phi^i\lvert_{\phi^i=v^i }  
  =    -ig^a A^a_\mu T^a\phi^i\lvert_{\phi^i=v^i }\\    
  &= - i\left(g^a \frac{c^a}{g^a} T^a A(x)_{\mu}\right)_{ji}v^i= - A(x)_{\mu}Q_{ji}v^i = 0\ ,    
 \end{align*}
 where $A(x)_{\mu}$ is  an arbitrary vector function of $x$, which can be identified with 
 the photon field. From Eq.~(\ref{eq:theorem1}) we get 
 \begin{align}
  A^a_{\mu}M^{ab}A^{b\mu}= 0\ ,
 \end{align}
showing that  $A^a_{\mu}=c^a A(x)_{\mu}$ is a null space vector of $M^{ab}$.
\subsection{Rank 2 tensors} 
For a rank 2 tensor the analysis is quite similar. 
The gauge transformation of a rank two tensor under 
the gauge group is 
\begin{align*}
 \Phi^{i'j'}= U^{i'}_{i} U^{j'}_{j}  \Phi^{ij}\ ,
\end{align*}
where the gauge group  transformation   $ U^{i'}_{i}(\theta(x))$ is a function of the
local coordinate $x$.
This allows us to define the covariant derivative as 
\begin{align*}
D_{\mu}\Phi^{ij}=
\partial_{\mu}\Phi^{ij}
 -ig^a\left(T^aA^a_{\mu}\right)^{i}_{\alpha}    \Phi^{\alpha j}
 -ig^a\left(T^aA^a_{\mu}\right)^{j}_{\alpha}   \Phi^{i\alpha } .
\end{align*}
For the $SU(3)$ gauge group,  $T^a = \frac{\lambda^a}{2}$, $g^a=g$,   $U(\theta)=\exp (-i\theta^a T^a)$ and
the gauge transformation of the vector field is
\begin{align*}
A^{\prime}_{\mu}=T^aA^{\prime a}_{\mu}= U(\theta)T^aA^{a}_{\mu}U(\theta)^{\dagger}+\frac{i}{g}U(\theta)\partial_{\mu}U^{\dagger}(\theta)\ . 
\end{align*}
We do not lose generality by assuming that the VEV of the  Higgs rank 2 tensor is the product of two Higgs 
scalars in the fundamental representation\footnote{
Any  component of a matrix  can always be written as the tensorial product of two vectors.}, \ie
$\Phi^{ij}=\chi^i\xi^j$. In similar way as we did for the rank 1 tensors 
we also build the null vector as  $A^a_{\mu}=\frac{c^a}{g^a} A(x)_{\mu}$, 
thus the covariant derivative is
\begin{align*}
D_\mu \Phi^{ij}\lvert_{\Phi=\langle\Phi\rangle}=& 
 -ig^a\left(T^aA^a_{\mu}\right)^{i}_{\alpha}  \chi^\alpha\xi^j 
 -ig^a\left(T^aA^a_{\mu}\right)^{j}_{\alpha}  \chi^i\xi^\alpha   ,\notag\\
 =&-iA(x)_{\mu}\left(Q^{i}_{\alpha}\chi^\alpha\xi^j+Q^{j}_{\alpha}\chi^i\xi^\alpha\right)     ,\notag\\
 =&-iA(x)_{\mu}\left(q^{i}\chi^i\xi^j+q^{j}\chi^i\xi^j\right)\\
 =&-iA(x)_{\mu}\left(q^{i}+q^{j}\right)\chi^i\xi^j\ , 
\end{align*}
where in the last step we take into account that the charge operator 
is diagonal, \ie $Q^i_{\alpha}\chi^{\alpha}=q^i\chi^{i}$ and $Q^j_{\alpha}\xi^{\alpha}=q^j\xi^{j}$.
In these expressions the $q_i$ are the charges of the components of a vector 
in the fundamental representation. 
If the component $\langle\Phi^{ij}\rangle $ correspond to the VEV of a Higgs field 
then $q^{i}+q^{j}=0$ and the kinetic Lagrangian becomes zero,
\begin{align*}
\mathcal{L}=\text{Tr}\left(
(D_\mu \Phi^{ij})^\dagger
 D_\mu \Phi^{ij}\right)\lvert_{\Phi=\langle\Phi\rangle}
 = \frac{1}{2}
 A^a_{\mu}M^{ab}A^{b\mu}=0\ .
\end{align*}
This shows that, as we already demonstrated for the rank 1 tensor, 
$A^a_{\mu}=\frac{c^a}{g^a} A(x)_{\mu}$ is a null vector of the 
 mass matrix $M^{ab}$. The procedure is similar  for an arbitrary tensor.  
\section{\texorpdfstring{$Z^{\prime}$}{zp1}\ \ \  couplings}
For the SM extended by a $U(1)^\prime$ extra factor, the neutral current interactions of the fermions are described by the Hamiltonian
\begin{align}\label{eq:hnc}
 H_{NC}=&  \sum_{i=1}^2 g_i Z_{i\mu}^0
 \sum_f\bar{f}\gamma^{\mu}\left(\epsilon_{\text{\bf L}}^{(i)}(f)P_\text{\bf L}+\epsilon_{\text{\bf R}}^{(i)}(f)P_\text{\bf R}\right)f\ ,
 \end{align}
where $Z^{0}_{1\mu}$ and $Z^{0}_{2\mu}$ are the weak basis states such that $Z^{0}_{1\mu}$ is identified with the neutral 
gauge boson of the SM,  $Z$,  and   $Z^{0}_{2\mu}$ with the $Z^{\prime}$;
the index $f$ runs over all the SM fermions in the low energy Neutral Current (NC) 
effective Hamiltonian $H_{NC}$, and $P_\text{\bf L}=(1-\gamma_5)/2$ and $P_\text{\bf R}=(1+\gamma_5)/2$. 
It is convenient to write Eq.~(\ref{eq:hnc}) in terms of the vector and axial charges
\begin{align}\label{eq:331hnc}
H_{NC} =&\frac{1}{2 }\sum_{i=1}^2  g_{i} Z_{i\mu}^0
 \sum_f\bar{f}\gamma^{\mu}\left(G_{V}^{(i)}(f)-G_{A}^{(i)}(f)\gamma_5\right)f\ ,
\end{align}
\noindent
where the chiral couplings $\epsilon_\text{\bf L}^{(i)}(f)$ and $\epsilon_\text{\bf R}^{(i)}(f)$ are linear combinations of 
the vector $G_{V}^{(i)}(f)$ and axial $G_{A}^{(i)}(f)$ charges given by $\epsilon_\text{\bf L}^{(i)}(f)=[G_{V}^{(i)}(f)+G_{A}^{(i)}(f)]/2$ 
and $\epsilon_\text{\bf R}^{(i)}(f)=[G_{V}^{(i)}(f)-G_{A}^{(i)}(f)]/2$.
The mass eigenstates $Z_{1\mu}$ and $Z_{1\mu}$  are given by
\begin{eqnarray*}
 Z_{1\mu}&=& \ \ Z^{0}_{1\mu}\cos\theta+Z^{0}_{2\mu}\sin\theta\ , \\
 Z_{2\mu}&=& -Z^{0}_{1\mu}\sin\theta+Z^{0}_{2\mu}\cos\theta\ .
\end{eqnarray*}
For the numerical calculations we use the expressions for the vector and axial charges shown  in the Appendices~\ref{appendixc1} and \ref{appendixc2},   
where most of the values in the Tables are being presented for the first time in the literature. We have also used
 $\sin^2\theta_W =0.231$ and $g_1\equiv g/\cos \theta_W = 0.743$.
\subsection{The 3-3-1 charges and  coupling strength}
\label{appendixc1}
\begin{table}[!htbp]
\begin{center}
\bgroup
\def\arraystretch{1.5}%
\scalebox{0.75}{
\begin{tabular}{|c|ccccc|c|ccccc|}
\hline
     &  \multicolumn{11}{|c|}{Chiral Charges }   \\
\hline  
                     &$l$           &  $e_{\text{R}}$ & $q$ &  $u_{\text{R}}$ & $d_{\text{R}}$ &    &$l$                   &  $e_{\text{R}}$                &$q$&$u_{\text{R}}$                 & $d_{\text{R}}$   \\
\hline                   
$\epsilon^{I_{R3}}$  & 0                    &-$\frac{1}{2}$&  0  &+$\frac{1}{2}$&-$\frac{1}{2}$ &$\epsilon^{I_{R8}}$&$\frac{-2}{2\sqrt{3}}$&$\frac{-1}{2\sqrt{3}}$&0  &$\frac{+1}{2\sqrt{3}}$&$\frac{+1}{2\sqrt{3}}$  \\
$\epsilon^{U_{R3}}$  &-$\frac{1}{2}$        &  0           &  0  &    0         &+$\frac{1}{2}$ &$\epsilon^{U_{R8}}$&$\frac{1}{2\sqrt{3}}$&$\frac{2}{2\sqrt{3}}$&0  &$\frac{-2}{2\sqrt{3}}$&$\frac{1}{2\sqrt{3}}$  \\
$\epsilon^{V_{R3}}$  &+$\frac{1}{2}$        &+$\frac{1}{2}$&  0  &-$\frac{1}{2}$&   0           &$\epsilon^{V_{R8}}$&$\frac{1}{2\sqrt{3}}$&$\frac{-1}{2\sqrt{3}}$ &0  &$\frac{1}{2\sqrt{3}}$&$\frac{-2}{2\sqrt{3}}$  \\
$\epsilon^{I_{BL}}$  &-$\frac{1}{2}$        &-$\frac{1}{2}$&+$\frac{1}{6}$&+$\frac{1}{6}$&$+\frac{1}{6}$&$\epsilon^{I_{L8}}$&-$\frac{1}{2\sqrt{3}}$&-$\frac{1}{\sqrt{3}}$&$\frac{1}{2\sqrt{3}}$&0             &0          \\
$\epsilon^{U_{BL}}$  &$0$                   &$0$           &+$\frac{1}{6}$&-$\frac{1}{3}$&$+\frac{1}{6}$   &$\epsilon^{V_{BL}}$& $0$                  &-$\frac{1}{2}$       &$\frac{1}{6}$        &$\frac{1}{6}$ &-$\frac{1}{3}$\\
\hline
\end{tabular}}
\caption{The chiral charges for the SM particles under the additional $U(1)$ symmetries embedded in  the $[SU(3)]^3$ group.
$l$ stands for the left handed doublet $(\nu_\text{L},e^-_\text{L})^T$ and $q$
for the quarks left handed doublet $(u_\text{L},d_\text{L})^T$.  For low energy constraints only the $Z'$ 
charges of the SM fermions  are involved in the calculation.}
\label{tabla:1}
\egroup 
\end{center}
\end{table}
\begin{table}
\begin{center}
\bgroup
\def\arraystretch{1.5}%
\scalebox{0.75}{
\begin{tabular}{|c|cccc|c|cccc|}
\hline
     &  \multicolumn{9}{|c|}{Vector and Axial Charges }      \\
\hline  
  &$u$ &  $d$ & $\nu$ &$e$ & &$u$ &  $d$ & $\nu$ &$e$    \\
  \hline
$g_V^{I_{R3}}$  & $\frac{1}{2}$         &$-\frac{1}{2}$         &$0$                    &$-\frac{1}{2}$         &      $g_V^{I_{R8}}$  & $\frac{1}{2\sqrt{3}}$ & $\frac{1}{2\sqrt{3}}$ & $\frac{-2}{2\sqrt{3}}$&$\frac{-3}{2\sqrt{3}}$ \\
$g_{A}^{I_{R3}}$& $-\frac{1}{2}$        &$\frac{1}{2}$          &$0$                    &$\frac{1}{2}$          &      $g_A^{I_{R8}}$  & $\frac{-1}{2\sqrt{3}}$& $\frac{-1}{2\sqrt{3}}$& $\frac{-2}{2\sqrt{3}}$&$\frac{-1}{2\sqrt{3}}$ \\
$g_{V}^{U_{R3}}$& $0$                   &$\frac{1}{2}$          &$-\frac{1}{2}$         &$-\frac{1}{2}$         &      $g_V^{U_{R8}}$  & $\frac{-2}{2\sqrt{3}}$& $\frac{1}{2\sqrt{3}}$ & $\frac{1}{2\sqrt{3}}$ &$\frac{3}{2\sqrt{3}}$  \\
$g_{A}^{U_{R3}}$& $0$                   &$-\frac{1}{2}$         &$-\frac{1}{2}$         &$-\frac{1}{2}$         &      $g_A^{U_{R8}}$  & $\frac{2}{2\sqrt{3}}$ & $\frac{-1}{2\sqrt{3}}$& $\frac{1}{2\sqrt{3}}$ &$\frac{-1}{2\sqrt{3}}$ \\
$g_{V}^{V_{R3}}$& $-\frac{1}{2}$        &$0$                    &$\frac{1}{2}$          &$1$                    &      $g_V^{V_{R8}}$  & $\frac{1}{2\sqrt{3}}$ & $\frac{-2}{2\sqrt{3}}$& $\frac{1}{2\sqrt{3}}$ &$0$                    \\
$g_{A}^{V_{R3}}$& $\frac{1}{2}$         &$0$                    &$\frac{1}{2}$          &$0$                    &      $g_A^{V_{R8}}$  & $\frac{-1}{2\sqrt{3}}$& $\frac{2}{2\sqrt{3}}$ & $\frac{1}{2\sqrt{3}}$ &$\frac{2}{2\sqrt{3}}$  \\
$g_V^{I_{BL}}$  & $\frac{1}{3}$         & $\frac{1}{3}$         & $-\frac{1}{2}$        & $-1$                  &      $g_V^{I_{L8}}$  & $\frac{1}{2\sqrt{3}}$ &$\frac{1}{2\sqrt{3}}$  &$\frac{-1}{2\sqrt{3}}$ &$\frac{-3}{2\sqrt{3}}$ \\
$g_A^{I_{BL}}$  & $0$                   & $0$                   & $-\frac{1}{2}$        & $0$                   &      $g_A^{I_{L8}}$  & $\frac{1}{2\sqrt{3}}$ &$\frac{1}{2\sqrt{3}}$  &$\frac{-1}{2\sqrt{3}}$ &$\frac{1}{2\sqrt{3}}$  \\
$g_V^{U_{BL}}$  & $-\frac{1}{6}$        &$+\frac{1}{3}$         &$0$                    &$0$                    &      $g_V^{V_{BL}}$  & $\frac{1}{3}$         & $-\frac{1}{6}$        & $0$                   & $-\frac{1}{2}$                            \\
$g_A^{U_{BL}}$  & $+\frac{1}{2}$        &$0$                    &$0$                    &$0$                    &      $g_A^{V_{BL}}$  & $0$                   & $\frac{1}{2}$         & $0$                   & $\frac{1}{2}$                             \\
 \hline 
\end{tabular}}
\caption{The vector and axial  charges for the SM particles 
under the additional $U(1)$ symmetries embedded in  the $[SU(3)]^3$ group. 
 For low energy constraints only the $Z'$ 
charges of the SM fermions  are involved in the calculation.}
\label{tab:gva}
\egroup 
\end{center}
\end{table}
For $X = U$,  $\cos \beta = d_U=-1$ and   Eq.~(\ref{eq:currents})  reduces to
\begin{align}
g_2J_{\mu2}=& -g_L J^I_{L8\mu}  \sin \alpha - g_R J^U_{R8\mu} \cos \alpha\ , 
\end{align}
where 
\begin{align*}
J^I_{L8\mu}=& \sum_i \bar{f}_i \gamma_\mu [\epsilon_\text{\bf L}^{I_{L8}}(i)P_L+\epsilon_\text{\bf R}^{I_{L8}}(i)P_R]f_i\ ,\\ 
J^U_{R8\mu}=& \sum_i \bar{f}_i \gamma_\mu [\epsilon_\text{\bf L}^{U_{R8}}(i)P_L+\epsilon_\text{\bf R}^{U_{R8}}(i)P_R]f_i\ , 
\end{align*}
In this way
\begin{align*}
g_2J_{\mu2}=& \frac{1}{2}\sum_i \bar{f}_i \gamma_\mu \bigg(-g_L \sin \alpha [g_\text{V}^{I_{L8}}(i) - g_\text{A}^{I_{L8}}(i)\gamma^{5}] \notag\\ 
                                                          &-g_R \cos \alpha [g_\text{V}^{U_{R8}}(i) - g_\text{A}^{U_{R8}}(i)\gamma^{5}]\bigg)f_i\ ,
\end{align*}
where
\begin{align}\label{eq:g's}
g_\text{V,A}^{X_{(L,R)8}}(i) =& \epsilon_\text{\bf L}^{X_{(L,R)8}}(i) \pm \epsilon_\text{\bf R}^{X_{(L,R)8}}(i)\ .
\end{align}
Reordering we have
\begin{align*}
g_2J_{\mu2}=& \frac{g_{331G}}{2}\sum_i \bar{f}_i \gamma_\mu \bigg(G_{V}^{331G}(i) - G_{A}^{331G}(i)\gamma^{5}\bigg)f_i,
\end{align*}
where the vector and axial charges  are
\begin{align*}
g_{331G}\,G_{V,A}^{331G}(i) = -g_L \sin \alpha \hspace{0.1cm}g_\text{V,A}^{I_{L8}}(i) - g_R \cos \alpha \hspace{0.1cm}g_\text{V,A}^{U_{R8}}(i)\ .
\end{align*}
In the differential cross-section always  appears 
the product $g_{331G}\,G^{331G}_{V,A}$,  where the $G^{331G}_{V,A}$ are 
the vector and axial charges in Eq.~(\ref{eq:331hnc}) and 
$g_{331G}$  is the corresponding coupling strength. 
For this reason, it is not necessary to know them  separately.  
Now, given that
\begin{align*}
g^{\prime} =& g_{L}\tan\theta_{W}\ ,\hspace{0.5cm}
g_{R} =
 \frac{2 g_{L} \sin\theta_{W}}{\sqrt{4\cos^{2}\theta_{W} - 1}}\ , \\ 
\cos\alpha =& 
\frac{g^{\prime}}{\sqrt{3}g_{L}} =
\frac{1}{\sqrt{3}}\tan\theta_W\ ,\hspace{0.5cm}
\end{align*}
we take the positive sign of $\sin \alpha$ in agreement with Eq.~(\ref{eq:eq331}) . The expressions for the vector and axial couplings  can be cast as
\begin{align}\label{eq:gv}
&g_{331G}\,G_{V,A}^{331G}(i) 
= \dfrac{-g_{L}}{\sqrt{3}\cos\theta_{W}\sqrt{4\cos^{2}\theta_{W} - 1}}\notag\\
&\times\bigg((4\sin^{2}\theta_{W} - 3)g_\text{V,A}^{I_{L8}}(i) + 2 \sin^{2}\theta_{W}g_\text{V,A}^{U_{R8}}(i)\bigg)
\ .
\end{align}
From Table~\ref{tab:e6charges} we obtain the chiral charges in Table~\ref{tabla:1} 
and their corresponding axial and vector expressions in Table~\ref{tab:331g}. By 
replacing these expressions in Eq.~(\ref{eq:gv}) we obtain 
the axial and vector charges as they are shown in Table~\ref{tab:331g}.
By defining $g_{331G}= g_{L}/\cos\theta_W$, as usual  for 3-3-1 models,  
we recover  the vector and axial couplings to the $Z'$ boson  in the G model~\cite{Salazar:2015gxa}. 
From Eq.~(\ref{eq:currents}), for $X=U$ and $X=V$ we obtain 
 exactly the same expression for the axial and vector  couplings 
 as the one for the $I$ case in Table~\ref{tab:331g}. 
 The reason behind of this coincidence is  that the  EW 
 Langrangians  $-\mathcal{L}^X= g_R J_{R3\mu}^{X}A_{R3}^{X\mu}+g_R J_{R8\mu}^{X}A_{R8}^{X\mu}$
 (see Eq.~\ref{eq:current}), are related each other
 by unitary transformations  for the different values of $X=I,U,V$,  as it is shown in Appendix~\ref{appendixa}.
 The same is not true for the left-right symmetric model and its  alternative models as we will see in the next Section. 
\begin{table}
\begin{center}
\begin{tabular}{|c|cc|}
\hline  
$f$             &$g_{331G}\,G_{V}^{\text{331}G}(f)$                                        &$g_{331G}\,G_{A}^{331G}(f)$ \\          
\hline
$\nu$  &$(\frac{1}{2}-\sin^{2}\theta_{W})\fac$             &$(\frac{1}{2}-\sin^{2}\theta_{W})\fac$ \\                
$e$    &$3(\frac{1}{2}-\sin^{2}\theta_{W})\fac$            &$(\sin^{2}\theta_{W}-\frac{1}{2})\fac$ \\
$u$    &$(\frac{4}{3}\sin^{2}\theta_{W}-\frac{1}{2})\fac$  &$-\frac{1}{2}\fac$  \\
$d$    &$(\frac{1}{3}\sin^{2}\theta_{W}-\frac{1}{2})\fac$  &$(\sin^{2}\theta_{W}-\frac{1}{2})\fac$ \\
\hline
\end{tabular}
\caption{Couplings for $Z_{\text{331G}}\rightarrow\overline{f}f$. Here $\fac =g_{331G}/\sqrt{4\cos^2\theta_W-1}$ and  $g_{331G}=g_1=g_L/\cos\theta_W $}
\label{tab:331g}
\end{center}
\end{table}
\begin{table}
\begin{center}
\begin{tabular}{|c|cc|}
\hline  
$f$             &$g_{IN}G_{V}^{IN}(f)$                                                    &$g_{IN}G_{A}^{IN}(f)\fact$    \\          
\hline
$\nu$  &$-\frac{1}{2}\fact$                                             &$-\frac{1}{2}\fact$ \\                
$e$    &$-\frac{1}{2}\fact$                                             &$-\frac{1}{2}\fact$ \\
$u$    &$0$                                                            &$0$  \\
$d$    &$\frac{1}{2}\fact$                                              &$-\frac{1}{2}\fact$ \\
\hline
\end{tabular}
\caption{Vector and axial  couplings $Z_{I}^{\text{Tri}}\rightarrow\overline{f}f$ ( $X = U$ case). Here  $\fact=g_R$.}
\label{tab:inertmodel}
\end{center}
\end{table}
The vector and axial  charges of  the $Z''$ current,  $g_2J_3$,   are obtained directly  from Eq.~(\ref{eq:currents}), 
\begin{align}
g_2J_2  = \quad\frac{g_{IN}}{2}\ \sum_i \bar{f}_i \gamma_\mu \bigg(G_{V}^{IN}(i) - G_{A}^{IN}(i)\gamma^{5}\bigg)f_i\ . 
\end{align}
Here we use $IN$ instead of  $I$  to denote the inert model $Z_I^{\text{Tri}}$ ,  in spite of the latter   is a more frequent label  for this model 
\footnote{That is  in order to avoid confusion with the label $I$ for the weak-$I$-spin symmetry.}. 
\subsection{Couplings for the left-right symmetric model and its alternative versions. }
\label{appendixc2}
From Eq.~(\ref{eq:lrcurrent}) the neutral current coupled to the $Z'$ boson is given by 
\begin{align}\label{eq:1}
g_2J_{2\mu} =& g_L \tan\theta_W \left(\alpha_X J_{R3\mu}^X-    \frac{c_XJ_{BL\mu}^X}{\alpha_X}\right),\hspace{0.3cm}X=I,V\ ,\notag\\
g_2J_{2\mu} =& g_L \tan\theta_W \left(\alpha_U J_{R8\mu}^U+    \frac{\sqrt{3}J_{BL\mu}^U}{\alpha_U}\right) ,\hspace{0.3cm}X=U
\end{align}
which  encompasses the three different X values.
From Eq.~(\ref{eq:1}) we get for  $X=I,V,U$  the  vector and axial charges for the left-right,  ALR   and inert models, respectively,
\begin{align*}
&g_2J_2  =  \frac{g_{(A)LR(U)}}{2}\\
&\sum_i \bar{f}_i \gamma_\mu \bigg(G_{V}^{(A)LR(U)}(i) - G_{A}^{(A)LR(U)}(i)\gamma^{5}\bigg)f_i\ ,
\end{align*}
where the index $(A)LR(U)$ stands for the three models, \ie $LR$, $ALR$ and $LRU$. 
\begin{align}\label{eq:4}
g_{(A)LR}G_{V,A}^{(A)LR}(i) &= g'\left(\frac{g_{V,A}^{X_{R3}}(i)}{\alpha_X}-c_X     \alpha_X g_{V,A}^{X_{BL}}(i)\right),\notag\\
g_{LRU}G_{V,A}^{LRU}(i)     &= g'\left(\frac{g_{V,A}^{U_{R8}}(i)}{\alpha_U}+\sqrt{3}\alpha_U g_{V,A}^{U_{BL}}(i)\right),
\end{align}
where, $g'=g_{L}\tan\theta_{W}$ and   $\alpha_I=\alpha_V=1/\sqrt{(4\cos^{2}\theta_{W}-1)}$. From  Table~\ref{tab:gva} and equations (\ref{eq:4}) we 
get the vector and axial-vector couplings to the $Z'$ boson, which are shown in Tables~\ref{tab:alrmodel} and \ref{tab:notlrmodel}. \\
\begin{table}
\begin{center}
\begin{tabular}{|c|cc|}
\hline  
$f$             &$g_{LR}G_{V}^{LR}(f)$                                                     &$g_{LR}G_{A}^{LR}(f)$ \\          
\hline
$\nu$  &$-\frac{1}{2}(1-4\cos^{2}\theta_{W})\facd$                      &$-\frac{1}{2}(1-4\cos^{2}\theta_{W})\facd$ \\                
$e$    &$(4\cos^{2}\theta_{W}-\frac{3}{2})\facd$                        &$\frac{1}{2}\facd$ \\
$u$    &$\frac{1}{3}(\frac{5}{2}-4\cos^{2}\theta_{W})\facd$             &$-\frac{1}{2}\facd$  \\
$d$    &$-\frac{1}{3}(\frac{1}{2}+4\cos^{2}\theta_{W})\facd$            &$\frac{1}{2}\facd$ \\
\hline
\end{tabular}
\caption{Vector and axial couplings for  $Z_{LR}^{\text{Tri}}\rightarrow\overline{f}f$ (The $X=I$ case). Here $\facd = g_{L}\tan\theta_{W}/\sqrt{4\cos^{2}\theta_{W}-1}$.}
\label{tab:lrmodel}
\end{center}
\end{table}
\begin{table}
\begin{center}
\begin{tabular}{|c|cc|}
\hline  
$f$             &$g_{ALR}G_{V}^{ALR}(f)$                                                     &$g_{ALR}G_{A}^{ALR}(f)$ \\          
\hline
$\nu$  &$\frac{1}{2}\facd$                                              &$\frac{1}{2}\facd$  \\                
$e$    &$(\frac{3}{2}-2\cos^{2}\theta_{W})\facd$                        &$\frac{1}{2}(4\cos^{2}\theta_{W}-1)\facd$ \\
$u$    &$\frac{1}{3}(4\cos^{2}\theta_{W}-\frac{5}{2})\facd$             &$\frac{1}{2}\facd$ \\
$d$    &$-\frac{1}{6}(4\cos^{2}\theta_{W}-1)\facd$                      &$\frac{1}{2}(4\cos^{2}\theta_{W}-1)\facd$ \\
\hline
\end{tabular}
\caption{Vector and axial  couplings for $Z_{ALR}^{\text{Tri}}\rightarrow\overline{f}f$ ( $X = V$ case). Here  $\facd = g_{L}\tan\theta_{W}/\sqrt{4\cos^{2}\theta_{W}-1}$.}
\label{tab:alrmodel}
\end{center}
\end{table}
\begin{table}
\begin{center}
\begin{tabular}{|c|cc|}
\hline\\[-14pt]  
$f$             &$g_{LRU}G_{V}^{LRU}(f)$                                                    &$g_{LRU}G_{A}^{LRU}(f)$    \\[3pt]          
\hline
$\nu_{\alpha}$  &$\frac{1}{2}\eta\alpha_U $                                          &$\frac{1}{2}\eta\alpha_U$ \\[3pt]                
$e_{\alpha}$    &$\frac{3}{2}\eta\alpha_U$                                          &$-\frac{1}{2}\eta\alpha_U$ \\[3pt]
$u_{\alpha}$    &$-\eta(\alpha_U + \frac{1}{2\alpha_U})$                     &$\eta(\alpha_U + \frac{3}{2\alpha_U})$  \\[3pt]
$d_{\alpha}$    &$-\frac{1}{2}\eta(\alpha_U + \frac{2}{\alpha_U})$                    &$-\frac{1}{2}\eta\alpha_U$ \\[3pt]
\hline
\end{tabular}
\caption{
Vector and axial  couplings for $Z_{LRU}\rightarrow\overline{f}f$ ( $X = U$ case)
Couplings $Z'\rightarrow\overline{f}f$ for $X = U$. Here  $\eta = g_{L}\tan\theta_{W}/\sqrt{3}$ and 
$\alpha_U = \sqrt{(g_R/g_L)^{2}\cot^2\theta_W - 3}$.}
\label{tab:notlrmodel}
\end{center}
\end{table}
\FloatBarrier
\bibliographystyle{apsrev4-1}
\bibliography{references331}


\end{document}